\documentclass[useAMS, usenatbib, 10pt]{mnras}

\usepackage{amsfonts}
\usepackage{amsmath}
\usepackage{amssymb}
\usepackage{color}
\usepackage{verbatim}
\usepackage{graphicx}
\usepackage{comment}
\usepackage{newtxtext,newtxmath}

\def \MSUN{\rm M_{\odot}}
\def \ZSUN{\rm Z_{\odot}}

\def \MTWOC{M_{\rm 200c}}

\def \MHALO{M_{\rm Halo}}
\def \MSTARS{M_{*}}
\def \MBH{M_{\rm BH}}
\def \KMS{{\rm km~ s}^{-1}}

\title[TNG50 Bubbles]{X-ray bubbles in the circumgalactic medium of TNG50 Milky Way- and M31-like galaxies: signposts of supermassive black hole activity}

\author[Pillepich et al.] {Annalisa Pillepich$^1$$^,$\thanks{E-mail: pillepich@mpia-hd.mpg.de}, Dylan Nelson$^{2}$, Nhut Truong$^{1}$, Rainer Weinberger$^{3}$, Ignacio Martin-Navarro$^{4,5}$, \newauthor Volker Springel$^{6}$, Sandy M. Faber$^{7}$, and Lars Hernquist$^{3}$
\\
$^{1}$Max-Planck-Institut f{\"u}r Astronomie, K{\"o}nigstuhl 17, 69117 Heidelberg, Germany\\
$^{2}$Universit\"{a}t Heidelberg, Zentrum f\"{u}r Astronomie, Institut f\"{u}r theoretische Astrophysik, Albert-Ueberle-Str. 2, 69120 Heidelberg, Germany\\
$^{3}$Center for Astrophysics | Harvard \& Smithsonian, 60 Garden Street, Cambridge, MA 02138, USA\\
$^{4}$Instituto de Astrofisica de Canarias, Calle Via Lactea s/n, 38200 La Laguna, Tenerife, Spain\\
$^{5}$Departamento de Astrofisica, Universidad de La Laguna, E-38205 La Laguna, Tenerife, Spain\\
$^{6}$Max-Planck-Institut für Astrophysik, Karl-Schwarzschild-Straße 1, 85740 Garching bei München, Germany\\ 
$^{7}$University of California Observatories/Lick Observatory,
University of California, Santa Cruz, CA, 95064}

\begin{document}
\maketitle
\begin{abstract}
The TNG50 cosmological simulation produces X-ray emitting bubbles, shells, and cavities in the circumgalactic gas above and below the stellar disks of Milky Way- and Andromeda-like galaxies with morphological features reminiscent of the eROSITA and Fermi bubbles in the Galaxy. Two-thirds of the 198 MW/M31 analogues inspected in TNG50 at $z=0$ show one or more large-scale, coherent features of over-pressurized gas that impinge into the gaseous halo. Some of the galaxies include a succession of bubbles or shells of increasing size, ranging from a few to many tens of kpc. These are prominent in gas pressure, X-ray emission and gas temperature, and often exhibit sharp boundaries with typical shock Mach numbers of $2-4$. The gas in the bubbles outflows with maximum ($95^{\rm th}$ pctl) radial velocities of $\sim100-1500~\KMS$. TNG50 bubbles expand with speeds as high as $1000-2000~\KMS$ (about $1-2$ kpc Myr$^{-1}$), but with a great diversity and with larger bubbles expanding at slower speeds. The bubble gas is at $10^{6.4-7.2}~$K temperatures and is enriched to metallicities of $0.5-2~\ZSUN$. In TNG50, the bubbles are a manifestation of episodic, kinetic, wind-like energy injections from the supermassive black holes at the galaxy centers that accrete at low Eddington ratios. According to TNG50, X-ray, and possibly $\gamma$-ray, bubbles similar to those observed in the Milky Way should be a frequent feature of disk-like galaxies prior to, or on the verge of, being quenched. They should be within the grasp of eROSITA in the local Universe.
\end{abstract}


\section{Introduction}

Over the last twenty years, a number of features observed towards the center of our Galaxy and extending for many kpc above and below its stellar disk have attracted significant attention. The discovery of a 200-pc wide bipolar structure at the Galactic center, which was detected in emission at mid-infrared wavelengths and that is consistent with a powerful nuclear starburst \citep{Bland-Hawthorn.2003},
followed that of a quasi-circular feature (the so-called North Polar Spur/Loop I) extending up to many tens of degrees above the Galactic plane and detected both in soft X-ray emission with ROSAT \citep{Egger.1994, Snowden.1995} and at radio wavelengths \citep{Berkhuijsen.1971}. Whereas the actual distance of the North Polar Spur remains debated to this date \citep[e.g.][]{Das.2020, LaRocca.2020}, the existence of nuclear activity at the Galactic center has been revealed by the discovery in {\it Fermi}-Lat data of the so-called Fermi bubbles \citep{Su.2010}: these are two large, $\gamma$-ray cocoons emerging from the Galactic center and extending above and below the Galaxy's disk up to about $8-10$ kpc in height and 6 kpc in width \citep{Selig.2015}.  

The Fermi bubbles are spatially correlated with the hard-spectrum microwave excess known as the WMAP haze \citep[at $23-60$ GHz,][]{Finkbeiner.2004}; their base appears connected to the Galactic center by the $140\times430$ pc shell features more recently detected with MeerKAT at 1.284 GHz \citep{Heywood.2019} and by the $300-500$ pc high X-ray chimneys discovered by \citealt{Ponti.2019}. The edges of the Fermi bubbles, at high galactic latitudes, also line up with features in the aforementioned ROSAT X-ray maps at $1.5-2$ keV. 

The interest in the Fermi bubbles, or more generally in the Milky Way (MW) bubbles, has been re-ignited by the discovery of the so-called eROSITA bubbles \citep{Predehl.2020}. The key finding from the new eROSITA data is the existence of a shell structure at $0.6-1.0$ keV extending {\it below} the Galactic plane and complementing the North Polar Spur. The two circular annuli encompass soft-X-ray-emitting bubbles that extend approximately 14 kpc above and below the Galactic plane and that exhibit remarkable morphological similarities with the Fermi bubbles. 

The origin of the Fermi bubbles has been debated extensively \citep[see][for a concise overview]{Yang.2018}. 
In particular, it remains uncertain whether these are the result of stellar feedback from the strong star formation (SF) occurring at the Galactic center \citep{Yusef-Zadeh.2009, Nogueras-Lara.2020} or of energy outbursts driven by the central supermassive black hole \citep[SMBH;][]{Genzel.2010}. Now, whether the $\gamma$-ray and the X-ray features are actually and physically related is an open question too, even though the most recent findings -- including the fact that the extended X-ray emission revealed by eROSITA coincides spatially with the soft component of the GeV emission reported to surround the Fermi bubbles \citep{Ackermann:2014} -- make it very plausible \citep{Predehl.2020}. Whatever the nature of the powerful energy injections at the Galactic center is, it is also still unclear whether the MW bubbles are the result of single or continuous events and whether, for example, the Fermi bubbles are driving the expansion of the eROSITA bubbles.

Constraints on the source of the energy injection(s) that inflated the MW bubbles may come from the kinematics of cold clouds in the regions of the bubbles. UV absorption lines in clouds about 10 kpc above and below the disk support the picture of gaseous outflows in the regions of the bubbles as fast as $900-1300~\KMS$ \citep{Fox.2015,Bordoloi.2017, Ashley.2020}. On the other hand, the kinematics of neutral hydrogen clouds just above and below the Galactic plane are consistent with radial velocities of 330 $\KMS$ \citep{DiTeodoro.2018}. Consequently, also the estimated expansion velocity and the age of the bubbles are highly uncertain, the latter ranging from 1 to 30 Myr \citep[][]{Fox.2015, Miller.2016}. 

Keeping in mind that the 'inferred physical properties' of the Milky Way bubbles are highly dependent on the underlying theoretical modelling and adopted assumptions, recent measurements of the linearly-polarized radio emission imply magnetic field intensities in the region of the bubbles of $6-16~\mu$Gauss \citep{Carretti.2013}. Moreover, the temperature of the eROSITA bubble(s) has been inferred to be about $3\times10^6$ K with Suzaku observations of X-ray emission \citep{Kataoka.2013} and $10^{6.6-6.7}$ K from the modeling of OVII and OVIII emission line strengths \citep{Miller.2016}. What is becoming progressively clear is the fact that the estimated energetics of the bubbles appear sufficient to modulate the structure and properties of  the circumgalactic medium (CGM) of the Milky Way \citep{Miller.2016, Predehl.2020}.

Numerical simulations of the Milky Way bubbles and associated features have been attempted over the last decade: some of their observed properties have been reproduced, and theoretical suggestions have been made for both a star formation (SF)- \citep{Sarkar.2015,Crocker.2015} and a SMBH-driven origin, the latter distinguished among ``jet'' \citep{Guo.2012,Yang.2012}, ``quasar outflow'' \citep{Zubovas.2012} and ``winds'' models \citep{Mou.2014}. In fact, specific observational aspects, such as the recent observations of the OVIII to OVII line ratios along sight lines passing through the Fermi bubbles \citep{Miller.2016}, have been shown to be recovered in numerical models with either SF or black hole accretion-driven wind mechanisms, the key aspect of the modeling being the amount, i.e. luminosity, of the energy injection \citep{Sarkar.2017}. However, so far most of the numerical experiments have been conducted by simulating one or the other possible driving mechanism, separately, and most of the work has focused on $\gamma$-ray observational signatures, i.e. on the Fermi bubbles, and hence on the emission and acceleration mechanisms of cosmic rays. Most importantly, the aforementioned works are based on idealized numerical experiments and, only at best, on 3D magnetohydrodynamical simulations of individual, isolated and idealized galactic gaseous haloes: hence, typically, no detailed modeling of the stellar and gaseous galactic disks nor the complex configurations of gaseous and dark matter (DM) haloes that result from the cosmological assembly of structures have been accounted for. 

In this paper, we overcome these limitations by analysing the output of the TNG50 simulation \citep{Nelson.2019, Pillepich.2019}: this is a gravity and magnetohydrodynamical run of a cosmological volume of about 50 Mpc on a side that has been developed to model the formation and evolution of thousands of galaxies in the full cosmological context. The numerical resolution may be coarser (yet still of the order of $50-300$ pc), and the implementation of feedback from SF and SMBH may be cruder, than in the aforementioned numerical experiments. However, the numerical scheme of TNG50 is designed to self-consistently realize the formation and evolution of galactic disks and of gaseous and DM haloes around them, and to follow the interaction of the energy injections from the central regions of the simulated galaxies with their interstellar and circumgalactic media. Importantly, the galaxies simulated within TNG50 turn out to be realistic, both from a population and demographic perspective as well as in terms of their sub-galactic and CGM physical properties (as detailed in Section~\ref{sec:TNG50}).

The starting point of our investigation stems from the fact that bi-polar outflows around simulated galaxies are an emergent phenomena of the stellar and AGN feedback processes in TNG50 \citep{Pillepich.2018, Nelson.2019}, given that both are implemented such that the energy is released isotropically at the injection scales. In particular, we have shown that feedback from SMBHs, particularly the low-accretion rate mode implemented as SMBH-driven winds \citep{Weinberger.2017}, is capable of driving $10^3~\KMS$ outflows that in turn carve low-density, over-pressurized, metal-enriched cocoon-like bubbles extending for tens of kpc above and below the disks of $z=1-2$ galaxies on their way to being quenched \citep{Nelson.2019}. This connection between isotropic energy injection and anisotropic outflow properties holds across a range of galaxy masses and cosmic times, within a cosmological numerical model. Analytical studies have also shown how isotropic feedback can explain bubbles and outflows propagating perpendicular to a galaxy plane, even in the case of quasar-like outflows and of the present-day Milky Way \citep[e.g.][]{Zubovas.2011}.

In this paper, we focus on $z=0$ Milky Way and Andromeda (MW/M31) -like galaxies in TNG50 and on the physical and observational manifestations of feedback in their CGM, with a particular focus on the thermodynamical, kinematics and metal-content properties of the gas. Our immediate goal is to derive connections to the recent X-ray observations of the Milky Way halo, i.e. to the eROSITA bubbles, and we postpone to future endeavours the task of forward modeling our simulated galaxies in $\gamma$-ray observables. 

The paper is organized as follows. In Section~\ref{sec:methods} we describe the TNG50 simulation, important aspects of the emerging TNG50 galaxy populations, the underlying numerical and physical model, and our selection of MW/M31-like galaxies. We give an overview of the pressure and X-ray morphologies of the halo gas in Section~\ref{sec:finding} and determine what types of galaxies preferentially exhibit bubbles in their CGM. We quantify the physical properties of the halo and bubble gas, the typical bubble sizes, outflow and expansion velocities, and estimated bubbles ages in Section~\ref{sec:properties}. We demonstrate in Section~\ref{sec:smbhfeedback} that, in our model, large-scale, coherent features of over-pressurized gas whose morphologies are reminiscent of the eROSITA and Fermi bubbles in the Galaxy are a manifestation of episodic kinetic energy injections driven by the SMBHs at the galaxy centers that accrete at low Eddington ratios. We discuss our results, implications, and limitations in Section~\ref{sec:discussion} and summarize our findings in Section~\ref{sec:conclusions}.

\begin{figure*}
		\includegraphics[width=14.0cm]{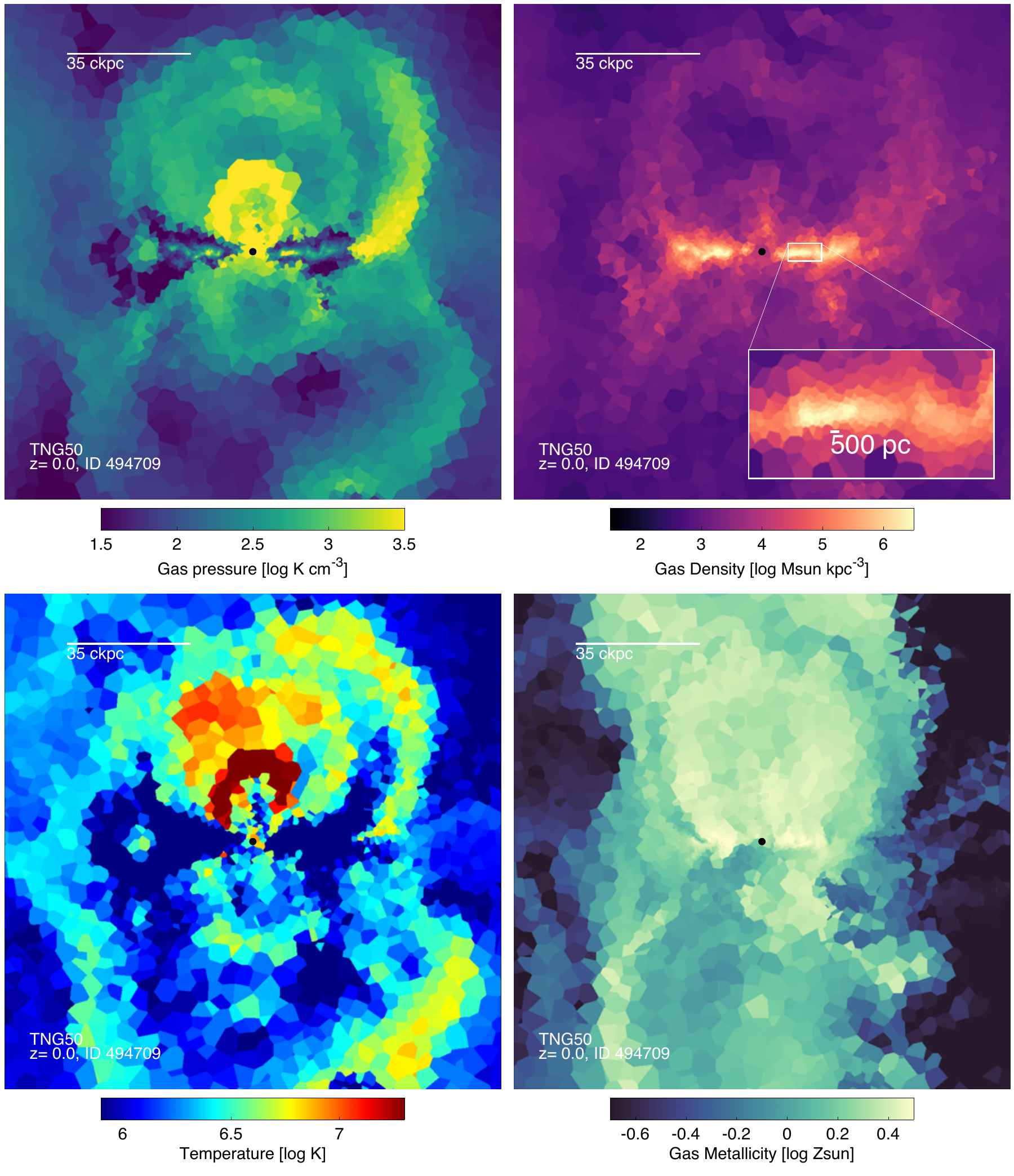} 
    \caption{A $10^{10.8}~\MSUN$ galaxy from TNG50 at $z=0$ viewed edge-on, across 140 kpc a side, and exhibiting bubbles and shells in the circumgalactic gas above and below its galactic plane. From the top left panel, clockwise, colors denote gas pressure, density, temperature and metallicity. The maps represent the cross-section of the Voronoi tessellation of the gas with a plane perpendicular to the galactic plane and passing through the galaxy center, here occupied by a SMBH: black circle. The quasi-Lagrangian nature of the {\sc arepo} code and the fixed target gas cell mass imply smaller gas cells at higher densities.}
    \label{fig:voronoi_maps}
\end{figure*}

\section{The TNG50 simulated galaxies}
\label{sec:methods}

\subsection{The TNG50 simulation}
\label{sec:TNG50}
The TNG50 simulation \citep{Nelson.2019, Pillepich.2019} is a cosmological simulation for the formation and evolution of galaxies across a periodic-boundary cube of 52 comoving Mpc a side. The equations for gravity and magneto-hydrodynamics in an expanding Universe are numerically solved with the code {\sc arepo} \citep{Springel.2010}, which combines a TreePM gravity solver with a quasi-Lagrangian, Voronoi- and moving-mesh based method for the fluid dynamics -- a visualization of the underlying Voronoi tessellation in a TNG50 galaxy is shown in Fig.~\ref{fig:voronoi_maps}. Prescriptions for gas cooling and heating, for star formation, stellar evolution, and metal enrichment, for feedback from supernoave and for seeding, growth and feedback of super massive black holes (SMBHs) are included in the calculations \citep{Weinberger.2017, Pillepich.2018}. 

In TNG50, a total of about 20 billion among dark-matter (DM), stellar, and SMBH particles and gas cells are simultaneously followed from $z=127$ to the current time of $z=0$, with individual time steps. This returns at recent epochs both a sample of the large-scale structure of the Universe as well as thousands of galaxies therein, simulated in different environments and at different evolutionary stages, with emerging sub-galactic features such as spiral arms, bulges, bars, and gas flows within and around them. 

TNG50 is the highest-resolution incarnation of the IllustrisTNG project \citep{Marinacci.2018, Naiman.2018, Springel.2018, Nelson.2018, Pillepich.2018b}, whose outcome has been studied and contrasted to observational findings across a wide range of applications: a partial summary is discussed in the public-release document \citep{Nelson.2019release}. Of relevance for the scientific messages of this paper, the IllustrisTNG model and TNG50 successfully return two distinct classes of galaxies -- star-forming and quiescent \citep{Pillepich.2019} --, with a population of quenched galaxies emerging also at intermediate and high redshifts \citep[$z\gtrsim2$,][]{Donnari.2020} and with the two populations separating in color \citep{Nelson.2018, Joshi.2021}, morphology \citep{Pillepich.2019, Joshi.2020} and kinematics \citep{Du.2021}. Importantly, both the global, e.g. stellar sizes, and small-scale morphological stellar features of IllustrisTNG galaxies have been shown to be in unprecedented agreement with observational results at low redshift \citep{Genel.2018, Rodriguez-Gomez.2019, Zanisi.2020} and, at least qualitatively, at intermediate redshifts \citep{Pillepich.2019}. 
The star-formation rate radial profiles of TNG50 galaxies are in the ball park of those of 3D-HST galaxies at $z\sim1$ \citep{Nelson.2021erica} and with those derived from MaNGA at $z\sim0$ \citep{Motwani.2020}; and the gas-phase metallicity gradients within TNG50 galaxies are consistent with those observed at $z=0-0.5$, although the preference for flat gradients observed in $z\gtrsim1$ galaxies is not present \citep{Hemler.2021}. Furthermore, the gaseous outflow properties around TNG50 galaxies can be very diverse \citep{Nelson.2019}, with outflows naturally exhibiting some collimation and bipolarity and with the average metallicity of the circumgalactic gas being higher along the minor axis of $\lesssim 10^{10.5}~\MSUN$ galaxies \citep{Peroux.2020}. Finally, the gaseous atmospheres within and around TNG50 galaxies are X-ray {\it brighter} for star-forming than for quiescent galaxies at the transitional mass scale of $10^{10-11}\MSUN$ in stellar mass, in tentative agreement with existing {\it Chandra} observations of nearby galaxies \citep{Truong.2020} and amid a significant abundance of small-scale, cold gas structures in the circumgalactic medium (CGM) of red, massive, elliptical galaxies \citep{Nelson.2020}. 

All the results listed above provide confidence on the underlying galaxy formation model of TNG50, including aspects of SMBH growth and feedback, including at the current epoch, and despite the necessarily incomplete and sub-grid nature of their physical modeling. In fact, in the TNG50 simulation, all this is achieved with a uniform numerical resolution of $8.5\times10^4~\MSUN$ for the target mass of gas cells and stellar particles and of $4.5\times10^5~\MSUN$ for the DM particle mass. The gravitational forces of stellar and DM particles are softened on scales smaller than 288 pc (at $z=0$), while the gravitational softening length of the gas cells is adaptive, fixed at 2.5 times the effective gas cell radius and with a minimum imposed at 72 physical pc. 
On the other hand, the quasi-Lagrangian nature of the {\sc arepo} code and the adaptivity of its Voronoi mesh mean that the magneto-hydrodynamics and the scales below which star formation and feedback are subgrid depend on local density, with smaller cells sampling higher densities (see Fig.~\ref{fig:voronoi_maps}) and with (de)refinement ensuring a fixed gas cell mass, within about a factor of two from the target mass. The average spatial resolution of the gas (i.e. cell size) in the star-forming regions of typical galaxies is about $100-200$ pc, and roughly $0.5-$some kpc within the CGM of $10^{12}\MSUN$ haloes. For reference, the smallest cell in the TNG50 box at $z=0$ spans 7 pc and within the star-forming regions across all TNG50 MW/M31-mass galaxies can be as small as 9 pc, but typically is about 60 pc. These spatial scales correspond to time steps as small as $0.02-0.05$ Myr (i.e. 50 kyr) in the innermost regions of galaxies, depending on the type of resolution element and of time step. Other details regarding the numerical resolution of TNG50 are given in \cite{Pillepich.2019, Nelson.2019, Nelson.2020}.

\subsection{The underlying galaxy physics model}
\label{sec:model}
The details of the galaxy formation and evolution ingredients of the IllustrisTNG simulations, and hence of TNG50, are provided and described in depth in \cite{Weinberger.2017, Pillepich.2018}. The model has been designed with the goal of producing realistic galaxies and galaxy populations across types, mass scales and cosmic epochs, i.e. not only to reproduce MW/M31-like objects. Here we succinctly recall the most relevant features that are needed to understand the findings of this paper.
Chiefly, in the IllustrisTNG model, feedback recipes from star-forming regions and from SMBHs are implemented in a subgrid and effective fashion so that mass, metals, energy, and/or momentum are exchanged between the evolving populations of stars and SMBHs, on the one side, and the interstellar (ISM) and circum/intergalactic medium (CGM/IGM) gas on the other. As a consequence, gas outflows develop across spatial and mass scales \citep{Nelson.2019}, regulate or even quench star formation \citep{Pillepich.2018, Weinberger.2017}, distribute mass and metals \citep[e.g.][]{Naiman.2018,Vogelsberger.2018, Torrey.2019}, and interact with the cosmological inflows of gas throughout the CGM and beyond. \\

{\it Star formation and stellar feedback.} In particular, in TNG50, gas cells are transformed into stellar particles if they exceed a fixed density threshold of $n_{\rm H}\sim0.1$cm$^{-3}$, the conversion being stochastic following the empirically defined Kennicutt-Schmidt relation and assuming a Chabrier initial stellar mass function. The realism of the ISM of TNG50 is limited in that the gas cannot cool below $10^4~$K and the relationship between temperature and density for the star-forming gas is determined by an effective equation of state to limit fragmentation and run-away collapse \citep{Springel.2003}. Stellar particles represent stellar populations that evolve and return mass and metals to the surrounding ISM via supernovae Type Ia (SNIa) and Type II (SNII) and asymptotic giant branch (AGB) stars according to look-up tables of mass and metal yields \citep[see][for details]{Pillepich.2018}. The rate and energy of SNII also determine the energy of stellar feedback: this is implemented in a so-called non-local fashion, whereby galactic scale outflows are launched directly from star-forming gas, with a prescribed wind velocity in random directions. In practice, wind particles are stochastically formed, spawned and hydrodynamically decoupled until they leave their surrounding ISM, as detailed in \cite{Vogelsberger.2013, Pillepich.2018}. Once below a certain surrounding density (see references above for details), wind particles hydrodynamically recouple to the gas and deposit their mass, momentum, metals, and thermal energy content. Due to the non-local nature of the stellar feedback in IllustrisTNG, the emerging properties of the gas outflows around galaxies must be interpreted with care within a few kpc distance from the injection regions, i.e. within and from the star-forming disks. On the other hand, despite the stellar feedback being injected isotropically, they can emerge in a bipolar manner by following the paths of least resistance, as demonstrated by \cite{Pillepich.2018, Nelson.2019, Peroux.2020}.\\

{\it SMBH seeding, growth, and feedback.} SMBHs with initial mass of $1.2\times10^6~\MSUN$ are seeded in haloes of friends-of-friends mass $\gtrsim7\times10^{10}~\MSUN$
and are implemented as sink particles, so that their mass can grow. As the simulation progresses, SMBHs can grow either because they accrete gas from the surrounding medium, with a prescribed Bondi–Hoyle–Lyttleton accretion rate capped at the Eddington limit, or because they merge with other SMBHs -- see \cite{Weinberger.2017} for all details. 

A fraction of the energy obtained via gas accretion is utilized to exercise feedback, which in turn naturally conditions the availability of gas for future SMBH mass growth. In the IllustrisTNG simulations, and so in TNG50 too, SMBH feedback comes in three flavors: thermal energy injection at {\it high} mass accretion rates, mechanical feedback at {\it low} accretion rates, and a third, radiative feedback channel also acting at high rates, whereby the cooling of the gas is modulated by the radiation field of nearby AGNs \citep[see][]{Vogelsberger.2013}. 

The separation between high vs. low, i.e. thermal vs. kinetic, modes of the IllustrisTNG SMBH feedback is implemented based on the following chosen prescription:
\begin{equation}
    \chi = \frac{\dot{M}_{\rm Bondi}}{\dot{M}_{\rm Edd}} =  
\begin{cases}
    0.1,                    & \text{if } \MBH \geq 7\times10^8\MSUN\\
    0.002(\MBH / 10^8)^{2},   & \text{otherwise.}
\end{cases}
\label{eq:chi}
\end{equation}
Namely, SMBHs with accretion rate larger (smaller) than the Bondi-to-Eddington mass growth ratio, $\chi$, are in thermal (kinetic) mode. Equation~\ref{eq:chi} implies that a SMBH can be a priori in either state, regardless of its mass, depending on its instantaneous accretion rate. Throughout this paper, high-accretion rates for TNG50 MW/M31-like galaxies correspond to Eddington ratios of $\gtrsim 0.5-1$ per cent, i.e. to mass accretion rates of a few $\times 10^{-3}\MSUN$ yr$^{-1}$ and above.

According to the subgrid implementation adopted in the fiducial IllustrisTNG model, in the high-accretion thermal mode, 2 per cent of the energy available from mass growth is donated continuously to the surrounding gas, in the so-called feedback region, in the form of internal energy: a thermal dump. For MW/M31-like galaxies at $z=0$, these choices correspond to thermal energy injections of about $10^{42-44}$ erg s$^{-1}$. In the low-accretion kinetic mode, up to 20 per cent of the SMBH energy is accumulated (see Eq.9 of \citealt{Weinberger.2017}) and then distributed in a pulsated manner to the gas cells around the black holes in the form of kinetic kicks. As we will detail in Section~\ref{sec:smbh_demographics}, these choices produce SMBHs in TNG50 MW/M31-like galaxies at $z=0$ that accrete at rates of $10^{-5}-10^{-3}\MSUN$ yr$^{-1}$ and hence store kinetic-feedback energy in the range $\lesssim 10^{41-43}$ erg s$^{-1}$, at most. The energy injections occur once a certain amount of energy has been stored, according to Eq.13 of \citealt{Weinberger.2017} and via the choice of a free parameter that influences the burstiness and thus the frequency of the kinetic kicks. In TNG50 MW/M31-like galaxies at $z=0$, the minimum injected energy per kinetic feedback event is typically a few $\times10^{56}~$erg. The individual gas cells within the SMBH-feedback region of TNG galaxies are hence kicked with velocities that depend on such available energy. If the available energy was uniformly distributed among the gas cells within the feedback region, each of them would be kicked with a velocity in the range of $600-1000$ km s$^{-1}$, averaging among gas cells and MW/M31-like galaxies. In fact, the kinetic energy is distributed among the gas cells in an SPH kernel-weighted manner, as per Eq.10 of \citealt{Weinberger.2017}, so that, for the typical TNG50 MW/M31-like galaxy, the gas cells within the inner part of the feedback region can receive kicks with velocities as high as $1200-11000$ km s$^{-1}$. In fact, these energy injections are administered to the gas around the SMBHs in random but different directions at each event, so that the feedback is isotropic at the injection scales when averaged across multiple events. We call `SMBH-driven winds' the emerging gas motions from this low-accretion rate feedback channel. No decoupling from the hydrodynamics is used for any SMBH-related feedback. 

It is important to note that the IllustrisTNG SMBH kinetic feedback mode is not meant nor designed to replicate the thin, collimated jets that have been observed in radio galaxies and that extend for tens of kpc into the gaseous haloes, which are then missing in our model; rather, it is theoretically motivated by the scenario of hot coronal winds from BH accretion flows \citep{Yuan.2014} and observationally connects to the recently observed `red-geyser' \citep{Cheung.2016, Roy.2018, Riffel.2019} and `FR0' \citep{Baldi.2016, Baldi.2019} galaxies. 
Moreover, it is not Equation~\ref{eq:chi} alone that determines when galaxies are in the thermal or kinetic feedback mode in the IllustrisTNG simulations: because of the combination of the hierarchical growth of structure together with the SMBH-feedback prescriptions described above and also with the chosen schemes and energetic of the {\it stellar feedback}, it turns out that in IllustrisTNG galaxies whose SMBH masses exceed $10^{8.1-8.2}~\MSUN$ at low redshifts are almost exclusively in low-accretion, kinetic feedback mode \citep{Terrazas.2020, Zinger.2020}. 
Finally, the feedback region over which SMBHs affect the gas is defined based on a prescribed number of neighbouring gas cells, in an SPH-like kernel-weighted fashion. In practice, in TNG50,
the radius within which 90 per cent of the SMBH feedback energy is injected is on average 480 pc for $\geq10^{10}~\MSUN$ galaxies at $z=0$ (median), but can vary between 350 pc and 3.7 kpc (10th-90th percentiles).

As detailed by \citealt{Pillepich.2018}, the choices for the SMBH physics described above have been adopted with the ultimate goal of reproducing simulated galaxy populations whose cosmic star formation rate density as a function of time and whose galaxy stellar mass function, SMBH mass vs. galaxy mass relation, halo gas fractions, stellar-to-halo mass relation and galaxy stellar sizes at $z=0$ are all in the ball park of available observational constraints (see e.g. their Figures 4 and 8). In fact, in addition to the results already summarized in Section~\ref{sec:TNG50}, a few analyses have focused specifically on the SMBH populations of $\gtrsim10^{10}~\MSUN$ galaxies produced by the IllustrisTNG simulations. Amid the enormous complexities of the simulation output vs. observations comparisons, acceptable agreement has been found with available observational results and expectations in terms of e.g. SMBH mass vs. galaxy mass relations at low redshift \citep{Weinberger.2018, Terrazas.2020, Habouzit.2021}; low-redshift ($z\lesssim1$, but not high-redshift) quasar bolometric \citep{Weinberger.2018} and hard X-ray [2–10 keV] luminosity functions \citep{Habouzit.2019}; fractions of AGNs and of obscured AGNs as a function of redshift \citep{Habouzit.2019}; and even correlations at $z\sim0$ between SMBH masses and X-ray temperature and luminosity of the surrounding gaseous haloes \citep{Truong.2021}.\\

{\it Effects of SMBH feedback, i.e. quenching.}
Within the IllustrisTNG model implementation -- similarly as within all other cosmological galaxy formation models that have been tested across large samples of high-mass galaxies --, no mechanism other than feedback from SMBHs has been shown so far to be capable of quenching entire populations of simulated massive galaxies as observations imply. In particular, within the IllustrisTNG model, it is the kinetic, SMBH-driven winds implementation of the mechanical feedback that is responsible for halting the star formation in massive ($\gtrsim$ a few $10^{10}\MSUN$ galaxies) galaxies \citep{Weinberger.2017, Terrazas.2020, Donnari.2020} and for making them red \citep{Nelson.2018, Donnari.2019}. The same SMBH feedback channel affects the thermodynamical and ionization states of the gas within and around galaxies \citep{Nelson.2018b, Truong.2020}. The SMBH-driven winds emerging in the IllustrisTNG simulations are both ejective -- i.e. they trigger quenching by removing gas from the star-forming regions of galaxies -- and preventative \citep{Zinger.2020} -- i.e. they heat up the gas, increasing its entropy and its cooling times also in the outer reaches of the halo, thereby preventing it from fuelling future star formation. Furthermore, the same low-accretion SMBH feedback limits the availability of gas for SMBH growth, so that IllustrisTNG SMBHs grow their mass predominantly either very rapidly while they exercise thermal mode feedback at low masses and high redshifts ($\MBH\lesssim 10^8~\MSUN$, i.e. $\MSTARS\lesssim 10^{10.5}~\MSUN$, i.e. $\MHALO\lesssim 10^{12}~\MSUN$) or more slowly via SMBH-SMBH mergers at higher masses and lower redshifts \citep{Weinberger.2018, Truong.2021}. In particular, at $z=0$, the average $10^8~\MSUN$ SMBH accretes gas with a rate lower than 0.1 per cent the Eddington rate \citep{Weinberger.2017}. As is the case for stellar feedback, although the feedback from SMBHs in the IllustrisTNG model is isotropic at the injection scales, the ensuing gaseous outflows on large scales are bipolar, with wide opening angles, directed along the galactic minor axes \citep{Nelson.2019}.\\

{\it Magnetic fields and shock finder.} Two noticeable features set the IllustrisTNG model apart in comparison to previous similar cosmological calculations. One is the inclusion of magnetic ﬁelds: in TNG50 ideal MHD is solved starting from an initial magnetic field seed of $10^{-14}$ comoving Gauss at $z=127$. Gas compression and shear motions due to the cosmological collapse of structures and of haloes and to the feedback from stars and SMBHs are responsible for amplifying the magnetic fields to the observed values of a few $\mu$G in and around galaxies \citep{Nelson.2018, Marinacci.2018}.

The other is the addition of a cosmological shock finder \citep{Schaal.2015}, so that shock surfaces can be identified and the dissipated energy rate and the Mach number of every gas cell can be measured: in TNG50 these are stored at each full snapshot. The shock finder will allow us to uncover interesting aspects of the interaction between SMBH-driven outflows and the gaseous atmospheres in and out of galaxies, in this and future papers.

\subsection{Sample selection: MW/M31-like galaxies from TNG50}
\label{sec:sample}
Within the volume simulated by TNG50, there are 898 galaxies at $z=0$ whose stellar mass (measured within 30 kpc) exceeds $10^{10}~\MSUN$, 565 of which are centrals of their dark-matter and gaseous halo. In this paper, we are interested in focusing on galaxies that are similar in mass, morphology, and environment to our Galaxy and Andromeda. We will expand to other types and mass ranges in future work.

Following \cite{Engler.2021} and \textcolor{blue}{Pillepich et al. in prep}, we select MW/M31-like galaxies as those TNG50, gravitationally-collapsed objects that have a stellar mass of $M_*(<30~{\mathrm{kpc}}) = 10^{10.5} - 10^{11.2}~\MSUN$ and that are disky in their stellar shape, i.e. have minor-to-major axis ratio of their 3D stellar mass distribution of $s < 0.45$ or appear disky by visual inspection of synthetic, 3-band stellar-light images in face-on and edge-on projections. We also require that no other massive galaxy with $M_* > 10^{10.5}~\MSUN$ is within a distance of $500~\rmn{kpc}$ of the MW/M31-like candidates and that the mass of the candidates' host halo is limited to $M_\rmn{200c}<10^{13}~\rmn{M}_\odot$, to avoid obvious members of massive groups.

This selection returns 198 MW/M31 analogues  at $z=0$, with median stellar mass of $5.4\times10^{10}\MSUN$ and median total host halo mass of $\MTWOC = 1.3\times10^{12}\MSUN$. Importantly for the scope of this paper, the majority of these galaxies are indeed late-type spiral galaxies, typically with thin gaseous disks as well: however, they can span a range of stellar disk lengths and heights, gas mass fractions and star formation rates, as well as of SMBH masses (see \textcolor{blue}{Pillepich et al. in prep}). For example, the median exponential stellar disk length of the selected sample is 4.4 kpc, with typical stellar thin-disk height of 450 pc. In fact, it is important to recall that the IllustrisTNG simulations have been developed to model galaxies across types, mass scales and cosmic epochs, and not tuned a priori to replicate in detail the global or inner properties of our own Galaxy or of Andromeda.

Most of the results presented in this paper are based on a population analysis of the galaxies in the aforementioned sample of MW/M31 analogues inspected at $z=0$. We will expand to other cosmic epochs at higher redshifts in future work. However, a handful of the selected galaxies can be found in the so-called subboxes \citep[see Section 3.1.4 of][]{Nelson.2019release}: these are fixed comoving sub-volumes within the TNG50 simulation domain for which the simulation output has been stored at much higher time cadence than the main snapshots. While the latter provide simulation data approximately every 150 Myrs, in the subboxes the time resolution of the data is from a few million years to about 8.5 million years. The analysis of galaxies in the subboxes allows us to follow their evolution in great detail: in this paper, we do so for selected galaxies and their progenitors where this is possible, for a time span of up to 1 billion years prior to $z=0$.

\begin{figure*}
		\includegraphics[trim=0 150 0 0,clip, width=18cm]{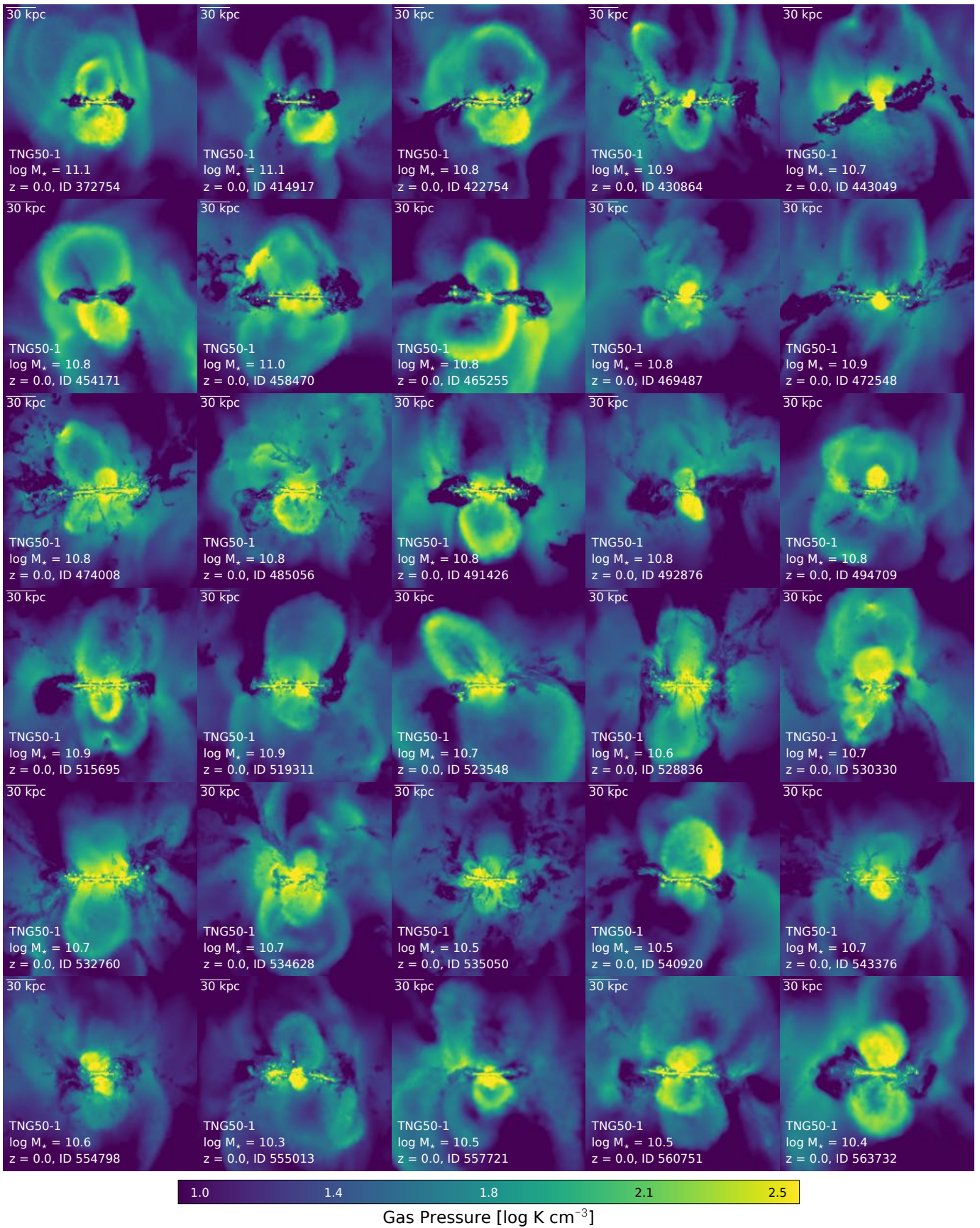}
    \caption{A selection of 30 galaxies among the 127 (of 198) MW/M31-like galaxies in TNG50 at $z=0$ that exhibit bubbles, in mass-weighted gas pressure. Each galaxy is shown edge-on, based on the total angular momentum of its stars and star-forming gas. The stamps are of fixed size (200 kpc per side, with a depth of 20 kpc) to highlight the diversity across the systems. Over-pressurized gas in spherical, bubble, dome-like, or shell configurations can be clearly identified, often extending both above and below the stellar disk within the same galaxy.} 
    \label{fig:top30_P_gas}
\end{figure*}
\begin{figure*}
		\includegraphics[trim=0 150 0 0,clip, width=18cm]{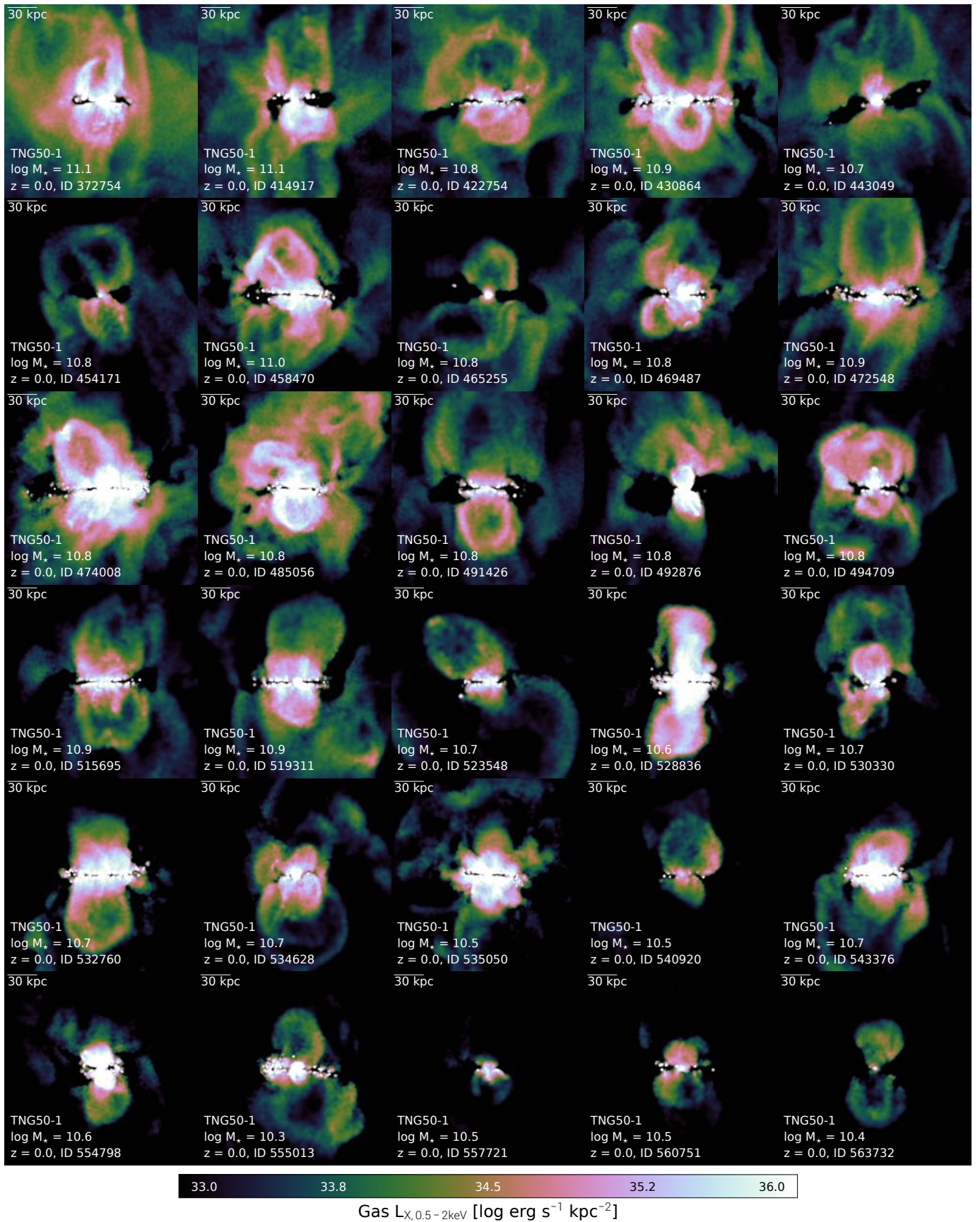}
    \caption{The same TNG50 galaxies as in Fig.~\ref{fig:top30_P_gas} but now shown in X-ray luminosity in the $0.5-2$ keV soft band. Systems of shells of gas are manifest in most galaxies, with the gas piling up along expanding fronts, producing under-luminous cavities in their wake. Some of the galaxies exhibit multiple shells of gas of increasing sizes that are also appreciable in X-rays.}
    \label{fig:top30_Xray}
\end{figure*}

\section{Bubbles in TNG50: diversity and galaxy demographics}
\label{sec:finding}

\subsection{Over-pressurized bubbles}
\label{sec:pressure}
TNG50 naturally returns bubbles and shells of over-pressurized gas above and below the disks of simulated galaxies. An example of such features can be seen in Fig.~\ref{fig:voronoi_maps}, for a TNG50 galaxy with stellar mass of $10^{10.8}~\MSUN$ at $z=0$ , i.e. very close to the stellar mass estimates of the Galaxy: the {\sc arepo} Voronoi tessellation of the gas is depicted as it is sliced by an imaginary plane passing through the galaxy's center and perpendicular to the galactic plane. The Voronoi cells are color-coded by gas pressure (left panel) and by the gas physical properties that determine their X-ray luminosity: gas density, temperature and metallicity. At high galactic latitudes, shells of compressed, high-density, high-temperature gas are clearly manifest in coherent, dome-like features in both directions above and below the galactic disk, with radii or heights ranging from about 20 to more than 70 kpc. 

By visually inspecting the gaseous content and properties of the 198 MW/M31 analogues from TNG50 (Section~\ref{sec:sample}), we can identify many cases whose morphological gaseous features are similar to those of Fig.~\ref{fig:voronoi_maps} and recall the ones seen in $\gamma$-rays and X-rays above and below the disk of our Galaxy (see Introduction).
In particular, we visually inspected edge-on maps of the simulated galaxies at $z=0$, spanning 200 kpc a side and 20 kpc in depth (i.e. across relatively thin layers), for the projected mass-weighted pressure of the gas. The edge-on projection is obtained by rotating each galaxy so that the vertical axis is aligned with the total angular momentum of the stars and gas in the galaxy (within twice the stellar half mass radius); the orientation of the horizontal axis of the edge-on view is aligned with the major stellar axis of each galaxy. We have examined a number of physical gas properties and conclude that maps of the gas pressure are the most promising for identifying galaxies whose gaseous features appear similar to those observed in the Galaxy. Our visual inspection in practice entails the search for over-pressurized coherent structures in the gas just outside the galaxy disk and all the way into the CGM.  

Roughly 2/3 (i.e. 127) of the MW/M31-like galaxies from TNG50 exhibit one or more large-scale, well-defined, dome-like or cocoon-like features of over-pressurized gas that impinge into the gaseous halo, extend above and/or below the stellar disk, and that stem from the galaxies' center. Additional examples of such bubbles and shells are shown in Fig.~\ref{fig:top30_P_gas}, for 30 random of such TNG50 galaxies seen edge-on. The quantity depicted in the maps is the mass-weighted gas pressure integrated along the line of sight: $P_{\rm gas}(x,y) = \int dz \, m_{\rm gas}(x,y,z) \, P_{\rm gas}(x,y,z) / \int dz \, m_{\rm gas}(x,y,z)$. The field of view is kept fixed to highlight the diversity across the systems, and the depth of the maps (i.e. the length of the line of sight in the integral above) is one tenth of the map extent, so 20 kpc. 
Most galaxies exhibit multiple features both above and below the stellar and gaseous disks; many galaxies display multiple shells in clear succession (e.g. ID 372754); some have hourglass, symmetric pressure distributions (e.g. ID 563732), while others have clearly asymmetric pairs of bubbles (e.g. 472548). Importantly, the over-pressurized gas encompasses regions that can span from a few kpc to at least $80-90$ kpc in heights -- we have focused on 200 kpc wide regions (see Section~\ref{sec:sizes} for more details).  

\begin{figure*}
\includegraphics[width=18cm]{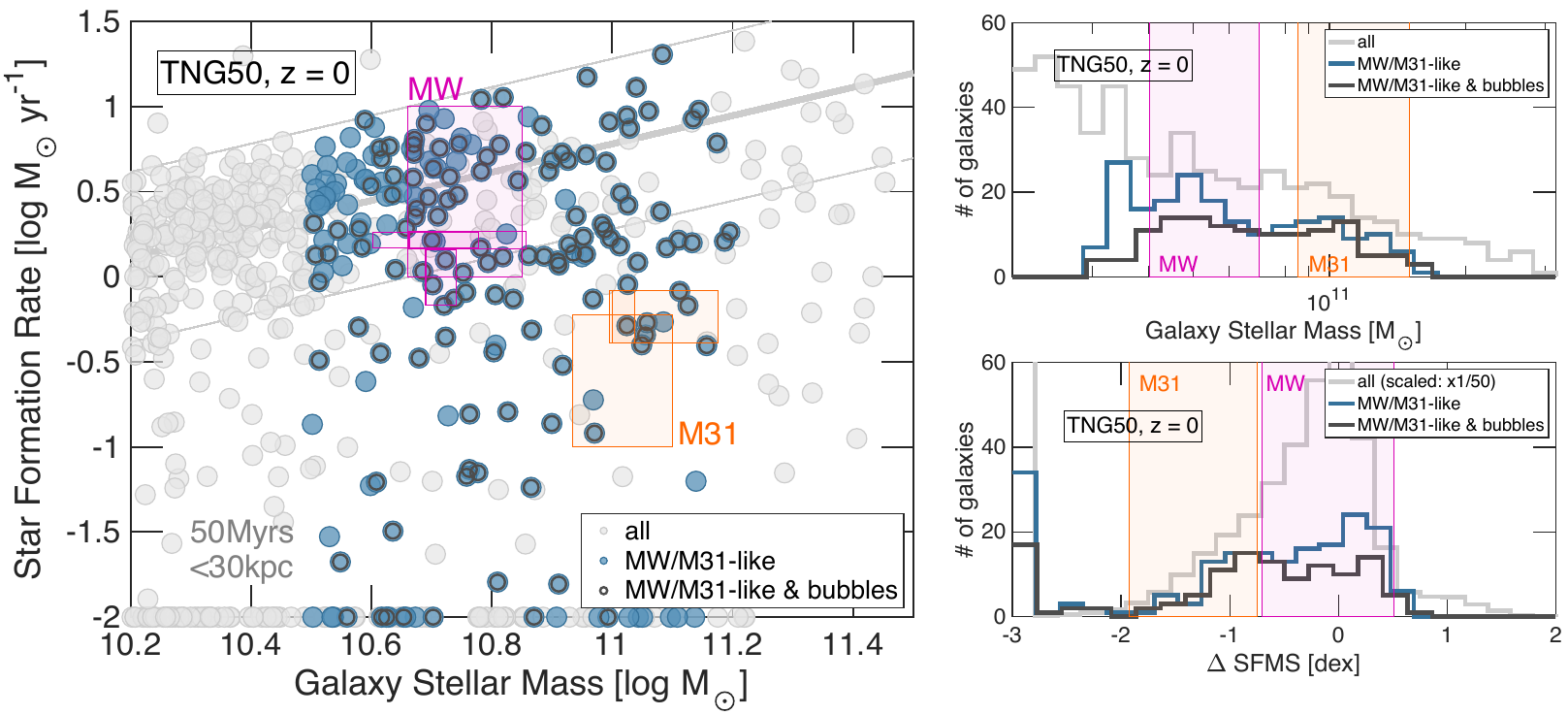}
    \caption{The distribution of TNG50 galaxies in the SFR-$\MSTARS$ plane (left) and as a function of stellar mass and logarithmic distance from the star-forming main sequence (SFMS, right panels). Light gray, blue and dark gray circles or curves indicate, respectively, all TNG50 galaxies {\it in the depicted ranges}, TNG50 MW/M31 analogues, and TNG50 MW/M31 analogues  that exhibit bubble features in their CGM. In the left panel, the SFMS of TNG50 galaxies is indicated with the solid thick (ridge) and thin gray lines (width of $\pm0.5$dex in specific SFR). Magenta and orange areas indicate current constraints from observations of the Galaxy and Andromeda. In TNG50, galaxies with bubbles are relatively less frequent at the low-mass end of the inspected sample and above or many dex below the SFMS.}
    \label{fig:demographics}
\end{figure*}

\subsection{X-ray emitting shells}

X-ray maps of the same galaxies trace quite faithfully, albeit in certain instances less prominently, the morphologies that are so evident in the gas pressure: see Fig.~\ref{fig:top30_Xray} for the X-ray emission in the 0.5-2 keV energy range. 

Here we forward model the simulated galaxies by determining the X-ray emission of each non star-forming gas cell given its simulated density, temperature, and metallicity and by adopting interpolated look-up {\sc XSPEC} tables \citep{Smith.2001} with a single temperature {\sc APEC} plasma model, following the methodology of \citealt{Truong.2020}. The integral of each contribution returns an intrinsic X-ray emission; no observational effects such as absorption along the line of sight or sensitivity, depth, and angular resolution of example observational campaigns are accounted for. 

As it can be seen in Fig.~\ref{fig:top30_Xray}, many of the depicted galaxies exhibit clear X-ray cavities above and below the stellar disk, contoured by shells of higher X-ray luminosities that reach $10^{35-36}$ erg s$^{-1}$ kpc$^{-2}$: these luminosities (corresponding to about $10^{-4}-10^{-3}$ cts s$^{-1}$ arcmin$^{-2}$) are within the grasp of eROSITA exposures of $25-200$ thousand seconds of extended sources within the local Universe, according to the estimates of \cite{Merloni.2012, Oppenheimer.2020}. Interestingly, some of the TNG50 bubbles show apparent X-ray morphologies similar to the eROSITA bubbles in the Galaxy, with higher luminosities in the proximity of the galactic planes and clearly visible under-luminous regions at higher galactic heights and enveloped by more luminous fronts, e.g. ID 532760, lower bubble.

In fact, the emissions estimated in Fig.~\ref{fig:top30_Xray} are somewhat of lower limits, by up to $0.5-1$ dex, as only the gas within a slice of 20 kpc along the line of sight contributes to the signal here, by construction. Namely, in such maps, the roughly spherical, hot gas component at larger distances is not accounted for, whereas for an external observer the latter would contribute with a roughly constant additive factor across the maps, but without any additional structural feature. We further expand on the spatial geometry of the X-ray luminosity of the bubbles and their detectability in Section~\ref{sec:x-rayprops}.

\subsection{Basic galaxy demographics}
\label{sec:demographics}

Over-pressurized and X-ray bubble features that resemble the eROSITA and Fermi bubbles in the Galaxy are, according to the TNG50 model, not rare. 
The top right panel of Fig.~\ref{fig:demographics} shows the basic demographics of the inspected sample (blue) and of the galaxies with bubble features (dark gray), as a function of galaxy stellar mass and in comparison to all TNG50 galaxies in the box (light gray histograms). As a notable difference with respect to the inspected MW/M31-like parent sample, there are relatively fewer galaxies with bubbles at the low-mass end of the distribution: this is shown in terms of galaxy stellar mass in top right panel of Fig.~\ref{fig:demographics} but holds also for total halo mass. Nonetheless, with TNG50 we have many tens of galaxies with bubbles at $z=0$ in the stellar mass range of $4\times10^{10}~\MSUN$ to $2\times10^{11}~\MSUN$, corresponding to total halo masses of $0.8-3\times10^{12}~\MSUN$. These are compatible with the mass estimates of the Galaxy and Andromeda, whose observational constraints are shown in magenta and orange shaded areas.

Fig.~\ref{fig:demographics}, left panel, shows the SFRs measured over 50 Myr and within a galactocentric aperture of 30 kpc for all TNG50 galaxies in the depicted mass range (gray full circles), for MW/M31 analogues  (blue full circles) and for MW/M31 analogues  that exhibit bubble features in their CGM (dark gray empty circles). The star forming main sequence (SFMS) simulated by TNG50 at $z=0$ is indicated with a gray thick line, and obtained via an iterative technique that progressively excludes quenched galaxies, i.e. galaxies that  fall 1 dex or more below it \citep[see][for details]{Pillepich.2019}.

MW/M31-like galaxies both with and without bubbles can be on the SFMS, in the green valley or can be quenched. Interestingly, also galaxies that are falling off from the SFMS can exhibit bubble features: however, within the MW/M31 mass range, galaxies with bubbles become increasingly rarer towards very low or very high levels of SF activity. This can be better appreciated in the lower right panel of Fig.~\ref{fig:demographics}, where the galaxies with bubbles are distributed based on their logarithmic distance from the SFMS at their stellar mass (dark gray histogram) in comparison to the MW/M31 sample (blue histogram): galaxies with bubbles are relatively less frequent both at the high and the very low SFR ends, namely they are less common in starbursts and in `fully' quenched galaxies than in main-sequence and green-valley systems.

We have checked (albeit do not show) that what distinguishes the dozens of galaxies with very low global SFRs and with over-pressurized bubble-like features in their CGM -- in comparison to those of similarly low SFRs but no coherent bubbles -- is the presence of a gaseous disk: the majority of the former are still characterized by a well-defined HI disk, even though with very low levels of star formation or H$\alpha$ gas. In comparison to the other galaxies of the sample, in particular those on the SFMS, they nevertheless lack HI column densities larger than $10^{19-20}$cm$^{-2}$ or total gas mass projected densities larger than $10^{6.5-7}~\MSUN$kpc$^{-2}$ in their disks. Upon closer inspection of this sample, it would also appear that the most-coherent, dome-like features in their CGM are very extended, many tens of kpc above or below the disks. However, a definitive assessment of their peculiarity in comparison to the rest of the galaxies remains elusive for now, given the low number statistics and the limitation of the visual inspection. 

A number of galaxies simulated within TNG50 can be found well within the current observational constraints on SFR and stellar mass for the Galaxy and Andromeda (magenta and orange boxes and shaded areas, respectively). Of those, 21 MW-like\footnote{Here we only count TNG50 simulated galaxies with bubbles that fall within the MW limits of Fig.~\ref{fig:demographics} and that are below the ridge of the SFMS: this choice is motivated by the fact that the majority of the observational constraints seem to favor a Galaxy whose SFR is below, rather than above, the SFMS.} and 9 M31-like analogues  also exhibit bubble features: this provides an even more tailored sub-sample of simulated galaxies for future inspection and for comparison to, or predictions of, observational signatures.

\begin{figure*}
		\includegraphics[width=17.5cm]{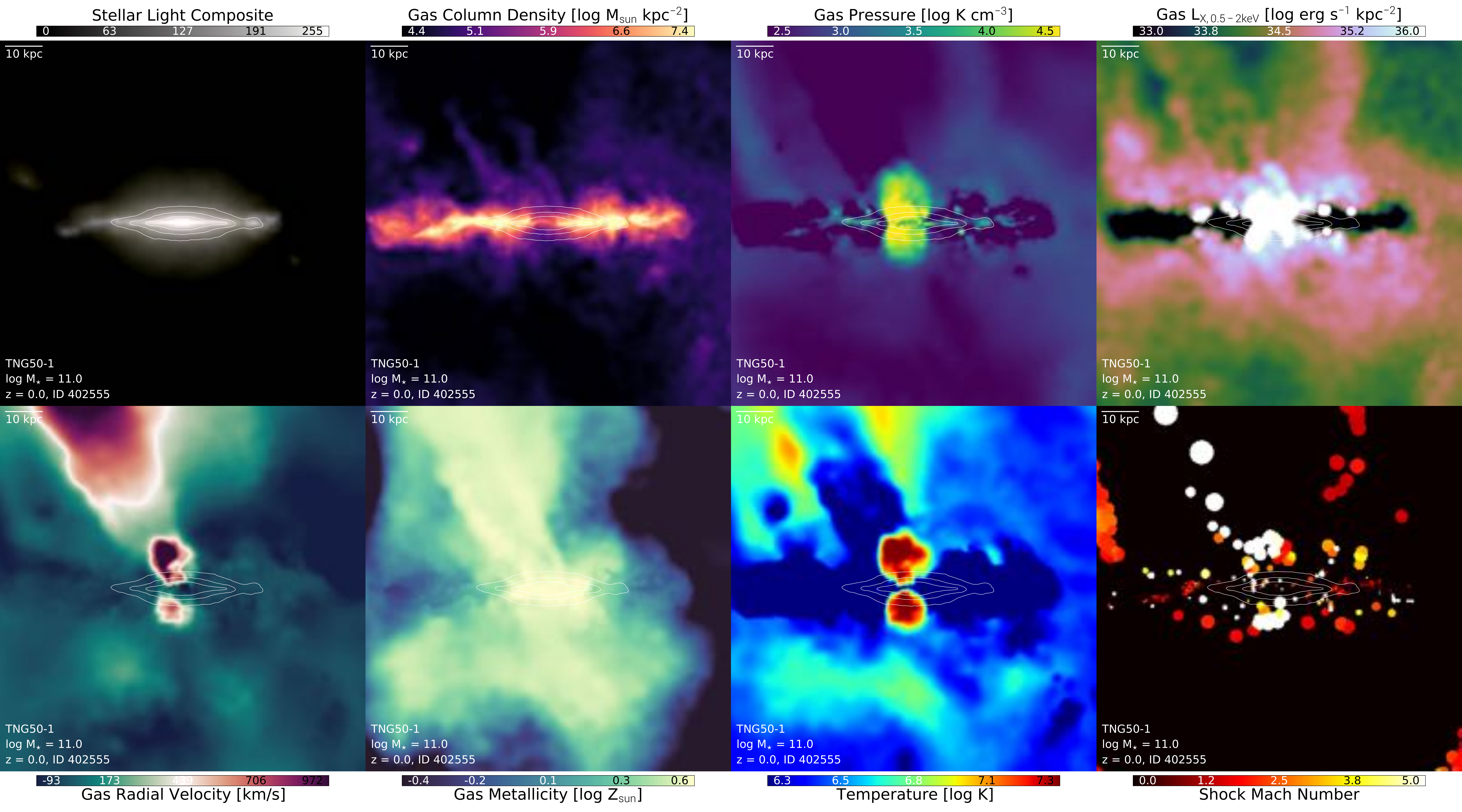}
    \caption{The properties of the gas within and around an example galaxy from the TNG50 simulation (Subhalo ID 402555) that exhibits two $\sim$10 kpc over-pressurized bubbles that are rather symmetric below and above the stellar disk. The panels depict from top left to bottom left (clock-wise): stellar light composite for the JWST NIRCam f200W, f115W, and F070W filters (rest-frame), gas column density, mass-weighted gas pressure, X-ray luminosity, mass-weighted Mach numbers in detected shocks, mass-weighted gas temperature, mass-weighted gas metallicity, and mass-weighted gas radial velocity. Each panel depicts the galaxy edge-on, according to its stellar and gaseous disk; the thickness of the slice is one tenth of the stamp extent, which is 100 kpc a side.}
    \label{fig:top2_1}
\end{figure*}

\begin{figure*}
		\includegraphics[width=17.5cm]{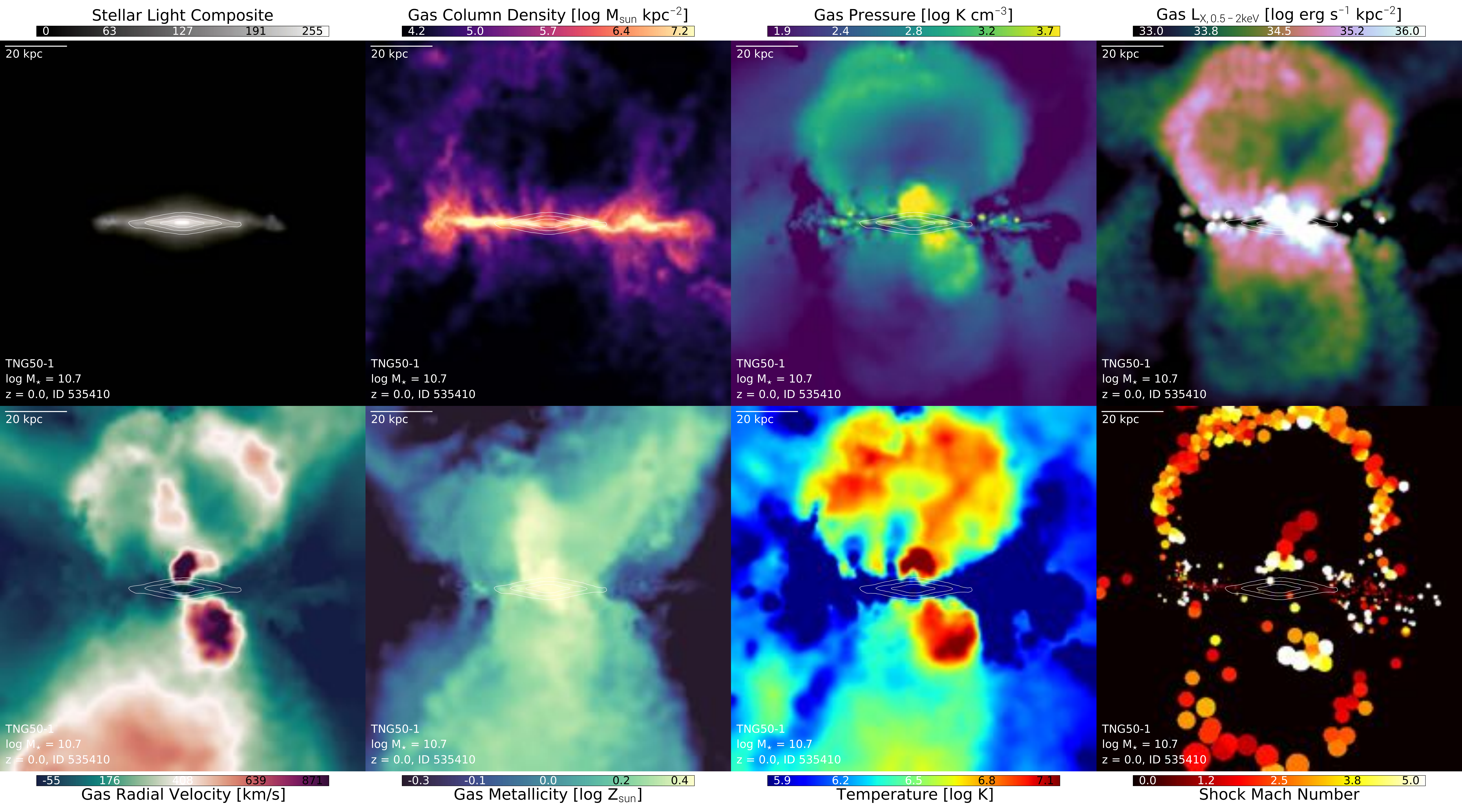}
    \caption{As in Fig.~\ref{fig:top2_1} but for an example TNG50 (Subhalo ID 535410) exhibiting an asymmetric succession of bubbles and shells, ranging from a few kpc to about 60 kpc in ``height'' and more. The stamps span 120 kpc a side. These maps support a push+shock mechanism, whereby high-velocity outflows from the innermost regions of the galaxies plough into the surrounding circumgalactic gas, often developing coherent shock fronts.}
    \label{fig:top2_2}
\end{figure*}

\section{Properties of the TNG50 bubbles}
\label{sec:properties}

\subsection{Physical properties of the gas}
Figs.~\ref{fig:top2_1} and \ref{fig:top2_2} showcase the gas properties within and around the bubbles of two example galaxies. The former exhibits two $\sim$10 kpc over-pressurized bubbles that are rather symmetric below and above the disk. The latter is an example of an asymmetric succession of bubbles and shells, ranging from a few kpc to about 60 kpc in ``height'' and more. The images depict slices whose line of sight depth is one tenth of the extent of the maps.

In both cases, the stellar disk of the galaxies is shown for reference in the upper left panels, in stellar light composite: both objects are clearly two good-looking disky galaxies. The corresponding gaseous disks are evident in the top, second from the left panels: gas column density also shows plumes or columns of gas more or less perpendicular to the disks, as well as cavities and accumulations of gas e.g. in a shell extending up to 60 kpc above the disk in Fig.~\ref{fig:top2_2}.

The base of the X-ray emission (top right panels) is larger for more extended, bubbles. The X-ray base appears wide even though the maps of the mass-weighted radial velocities of the gas (bottom left panels) clearly suggest an origin of the outflows that is very close to the galaxy center. In the galaxies of Figs.~\ref{fig:top2_1} and \ref{fig:top2_2}, the gas is outflowing with radial velocities as large as $\sim 800-1000$ km s$^{-1}$ (see next Sections for a more detailed quantification of the gas and bubble velocities). However, the geometry of the outflow velocities is complex: the gas in the middle of the most extended bubbles can move faster than the gas that accumulates at the edges; on the other hand, for the smaller dome-like features, the gas within them appears to move at maximal speeds. The metallicity maps (bottom panels, second from the left) show wide but prominent columns of metal-enriched gas (here up to around $2-4~\ZSUN$) co-spatial with the fast outflows and directed perpendicularly to the stellar and gaseous disks of the galaxies.

Not surprisingly, the mass-weighted temperature of the gas (bottom panels, second from the right) exhibits similar geometries and patterns as the gas pressure and X-ray emission. The bubble gas is at temperatures as high as $10^{6.4-7.2}$K. This is the case not only for the two example galaxies of Figs.~\ref{fig:top2_1} and \ref{fig:top2_2}, but more generally: maps of the gas temperature of the 30 galaxies of Fig.~\ref{fig:top30_P_gas} are given in the Appendix, in Fig.~\ref{fig:top30_T}. The bubble gas is clearly heated up by some internal i.e. galactic mechanism to temperatures that are up to one order of magnitude higher than the typical virial temperatures of the hosting haloes -- the virial temperature of a $10^{12}~\MSUN$ halo being $10^{5.8}$ K. The temperature in the TNG50 bubbles gas is in the ballpark of the estimated temperature of the eROSITA bubble(s), which reads about $10^{6.5}$ K \citep[see Introduction and ][]{Kataoka.2013}, and of the bubble/shell modeling of the OVII and OVIII emission line strengths of the Fermi bubbles: $10^{6.6-6.7}$ K \citep{Miller.2016}.




Finally, the high-velocity outflows produce shocks, sometimes with coherent shock fronts at the edges of the dome-like features that have been clearly identified in gas pressure, X-ray luminosity and temperature: thanks to the shock finder, we can record the Mach numbers of the shocks, as shown in the lower right panels of Figs.~\ref{fig:top2_1} and \ref{fig:top2_2} for mass-weighted average Mach numbers of gas cells undergoing a shock, i.e. only considering gas cells with recorded Mach number $\geq1$. Additional Mach number maps of TNG50 galaxies are given in the Appendix, in Fig.~\ref{fig:top30_machnum}. Coherent shock fronts can be clearly seen in many TNG50 galaxies, particularly for bubbles that extend to high galactocentric heights, of some tens of kpc. The Mach numbers of the gas in these shock fronts can be as high as a few, and we expand on their statistics in the next Sections. However, a more detailed analysis and possibly higher-resolution re-simulations would be needed to determine whether, or in what physical circumstances, the transition layers of Fig.~\ref{fig:top30_machnum} represent termination shocks \citep{Lacki.2014}, forward shocks \citep{Fujita.2013} and/or contact discontinuities \citep{Crocker.2011, Guo.2012, Mou.2014, Sarkar.2015} and whether multiple shocks and reversed shocks are in place. Based on the maps, it would appear that both density and (thermal) pressure jump, and so we call them generally ``shocks''; however, whether there is effectively particle transport through the layers and what the magnetic fields do will be determined in future analyses.

Overall, the diagnostics outlined so far support a push+shock, rather than a jet+shock \citep[e.g.][]{Zhang.2020}, mechanism for the development of the large-scale bubbles in TNG50, in that the geometry of the small-scale gas flows does not need to be collimated. In fact, the only source of energy in the IllustrisTNG simulated galaxies that could be responsible for pushing the gas are feedback from supernova and feedback from the SMBHs: we demonstrate which one of these two is the culprit in Section~\ref{sec:smbhfeedback}. In any case, in TNG50, the phenomena we are describing here are not the result of a jet+shock scenario, as in the TNG50 simulation no feedback mechanism is implemented to produce sharply-collimated jets but all feedback mechanisms are instead implemented as ``winds'' (see Section~\ref{sec:model} for details). Irrespective of the precise small-scale geometry, however, the gas in the innermost regions of galaxies is pushed outwards with velocities as high as many hundreds, if not thousands, $\KMS$, it hits the surrounding medium and heats it via shock dissipation, in agreement with the picture quantified via idealized experiments of the IllustrisTNG model by \cite{Weinberger.2017}. 

\begin{figure*}
		\includegraphics[width=17cm]{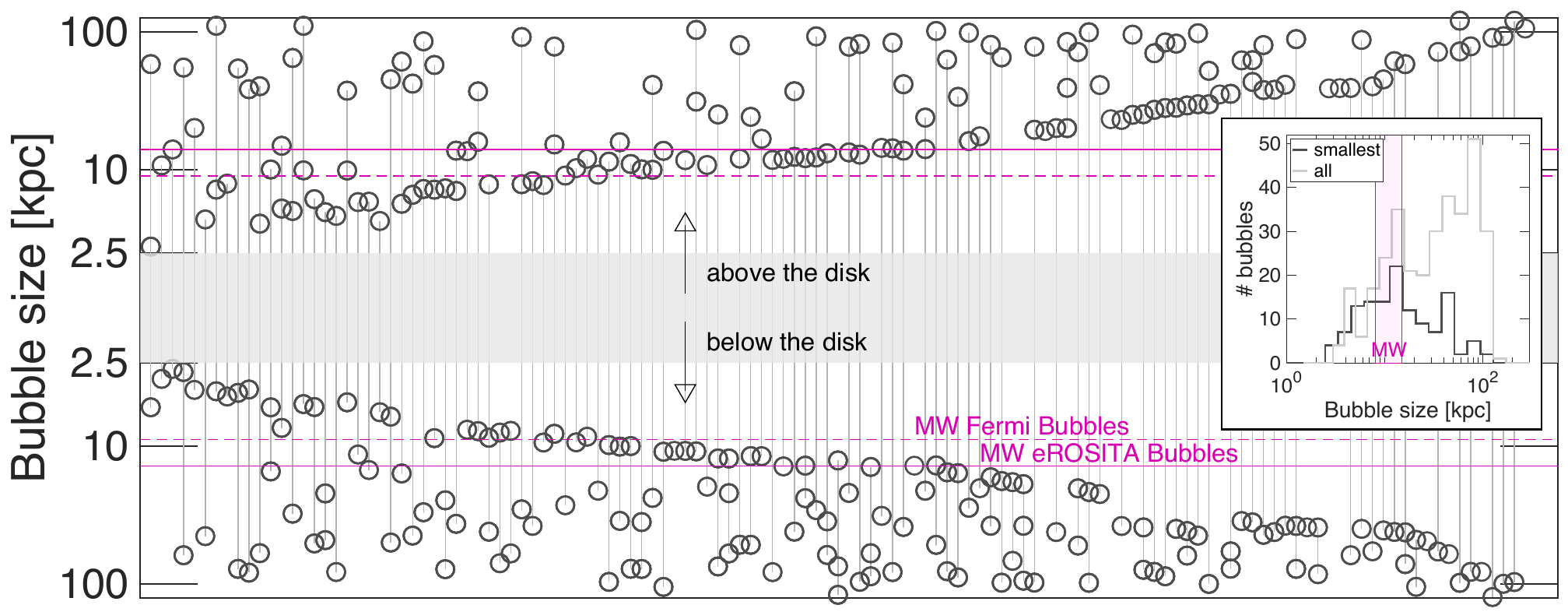}
    \caption{Physical extent of the bubbles for the 127 TNG50 galaxies among the parent sample of 198 MW/M31 analogues that exhibit bubbles or shells in the gas pressure, either above (positive) or below (negative values) the stellar disk, with the vertical axis aligned with the angular momentum of the stars and star-forming gas. Each galaxy is represented by one column, with one or more bubbles each; all galaxies are ranked from left to right based on the size of their smallest bubble; the y-axis scaling is logarithmic. Here by size we mean either the height or the maximum extent in the case of features that are inclined with respect to the minor axis of the galaxy. The magenta lines indicate the estimated heights for the X-ray (eROSITA) and $\gamma$-ray (Fermi) bubbles of the Galaxy, respectively at $\pm14$ and $\pm9$ kpc. The gray shaded area excludes $\pm2.5$ kpc where the bubbles would be too small to be identified. The inset shows the distributions of the smallest bubble in each galaxy (black) and all the bubbles in the sample (gray histogram). In TNG50, bubble sizes can range from a few kpc to at least about 100 kpc.}
    \label{fig:sizes}
\end{figure*}

\subsection{Heights or physical sizes}
\label{sec:sizes}

As appreciable in the Figures introduced so far, the extents of the bubbles identified in TNG50 galaxies at $z=0$ can be very diverse, ranging from a few kpc to tens of kpc. 

Fig.~\ref{fig:sizes} reports the distribution of the height or maximal size in kpc of TNG50 bubbles, the latter estimate being used in the case of features that are not perpendicular to the galactic disks. For each galaxy, we record and report in Fig.~\ref{fig:sizes} up to 4 features, and hence up to 4 bubble sizes, per galaxy. The galaxies are sorted based on the size of their smallest bubble, from small to large. 
Magenta lines and annotations indicate the estimated heights of the eROSITA and Fermi bubbles of the Galaxy, seen respectively in X-rays and $\gamma$ rays and extending up to about 14 and 9 kpc above and below the disk, respectively. In TNG50, there are 45 galaxies with at least one bubble exhibiting an extent similar to that of the Galaxy's bubbles (range of $8-14$ kpc); 12 of these galaxies have each a pair of bubbles with similar size and in the aforementioned range.

For the purposes of this Figure and the next steps of the analysis, the sizes of the bubbles are determined visually, by recording the position of the pressure edges in the gas pressure maps in the direction perpendicular to the galactic disks. The edge-on projections are determined as in Figs.~\ref{fig:top30_P_gas} and \ref{fig:top30_Xray}, with the minor axis aligned to the angular momentum of the stars and star-forming gas. In the case of inclined dome-like features in the inspected edge-on projection (i.e. cocoons that do not extend perpendicularly to the disk), we take the approximate locus of the pressure edges along the direction that maximizes the bubble extent. We have inspected maps only in one projection and spanning up to 100 kpc above and below the galactic disks: therefore, while there may be simulated bubbles that extend even further than 100 kpc, we do not keep track of those in this paper. 

While we are aware that an automatic procedure may be more desirable, we postpone to future work the task of developing an automated method to both identify bubble-like features as well as to extract their physical properties. 
Still, the visual inspection is sufficient for the task at hand, with the caveats that the bubble sizes should be taken with an uncertainty of a few kpc and that, by construction, we cannot identify bubbles smaller than $\lesssim2-3$ kpc, because of the confusing presence of the gaseous disks.

\subsection{Outflow velocities and bubbles expansion}
\begin{figure*}
		\includegraphics[width=8cm]{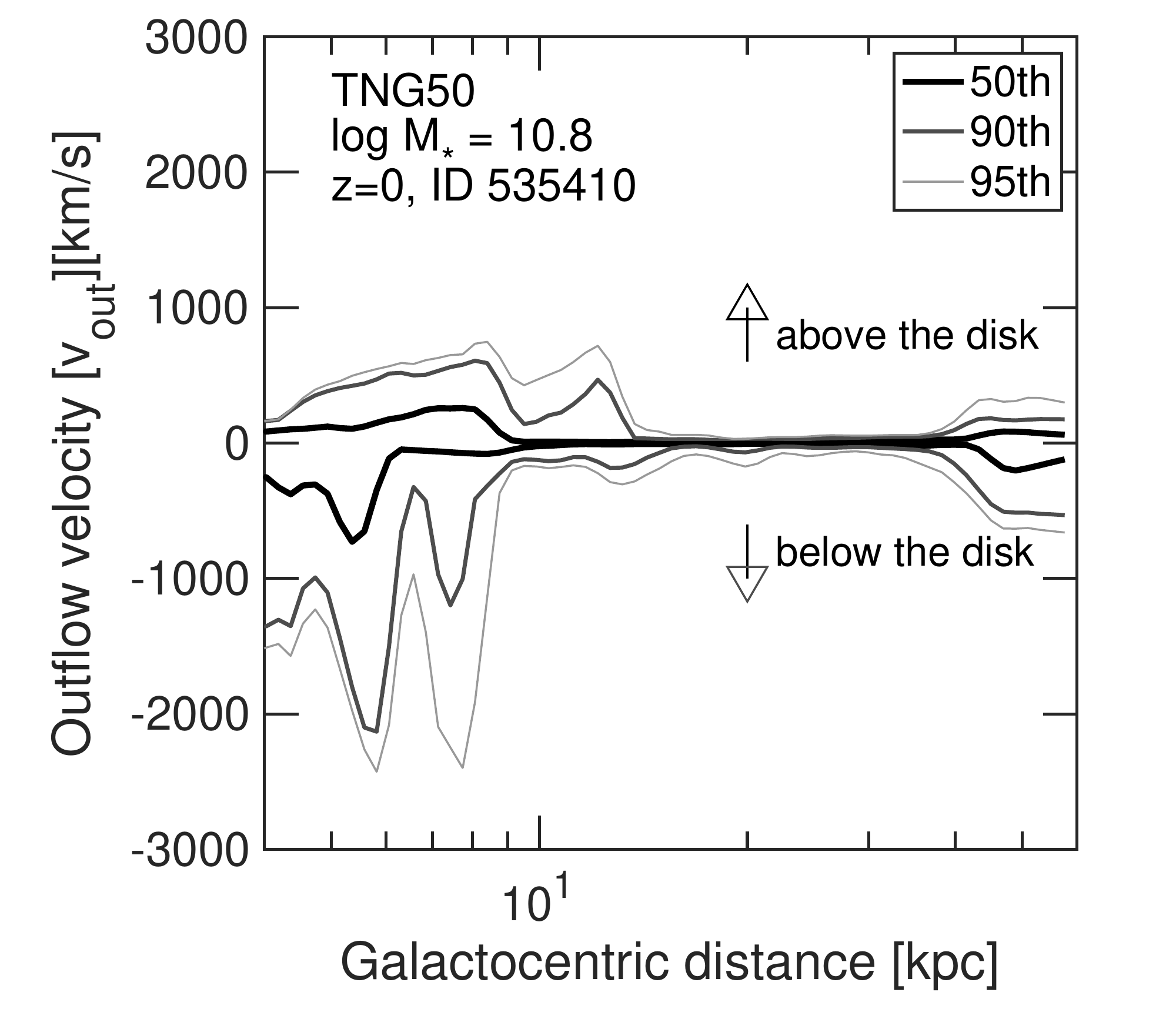}
		\includegraphics[trim=50 0 0 0 ,width=6.75cm]{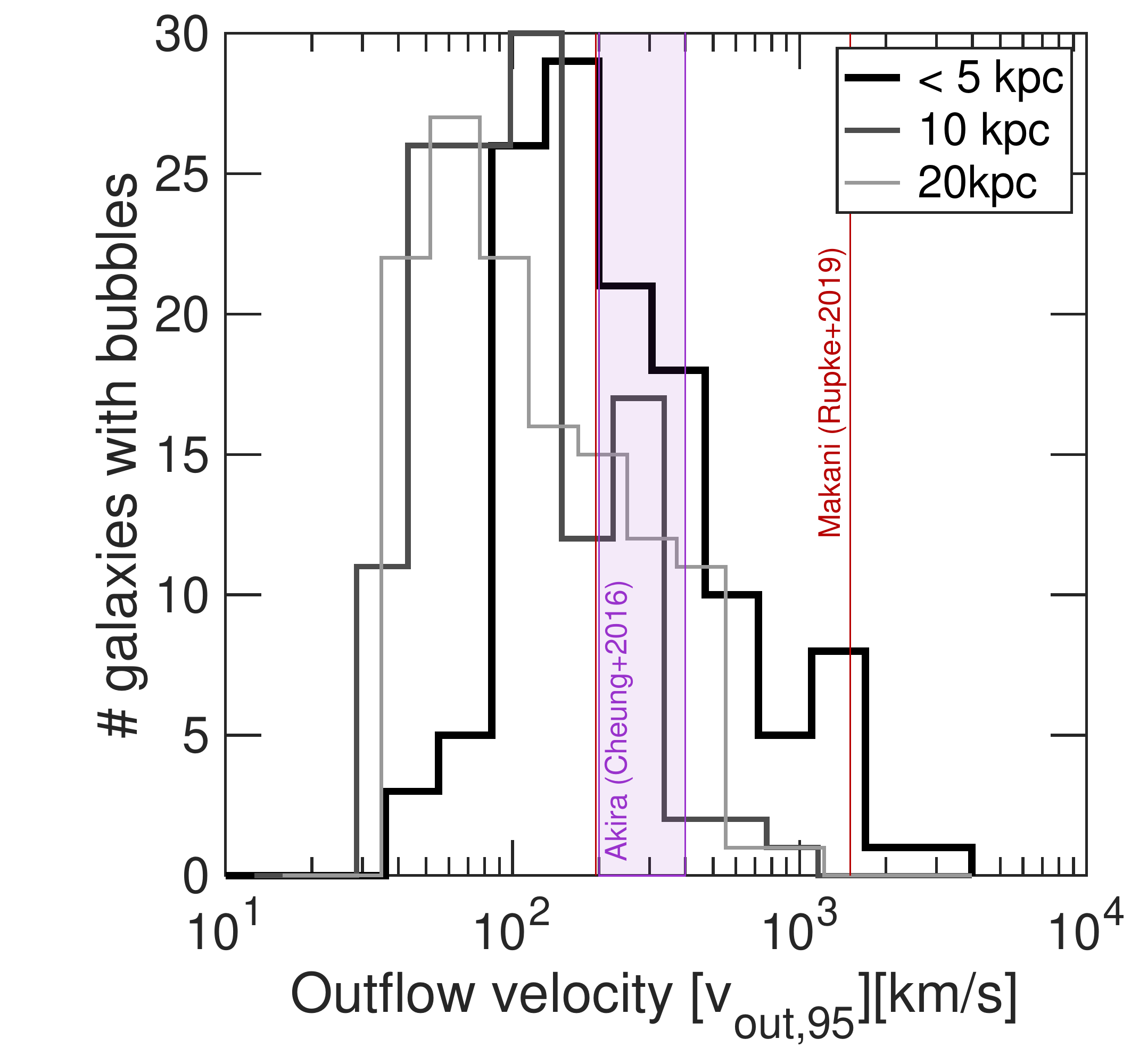}
		\includegraphics[width=16cm]{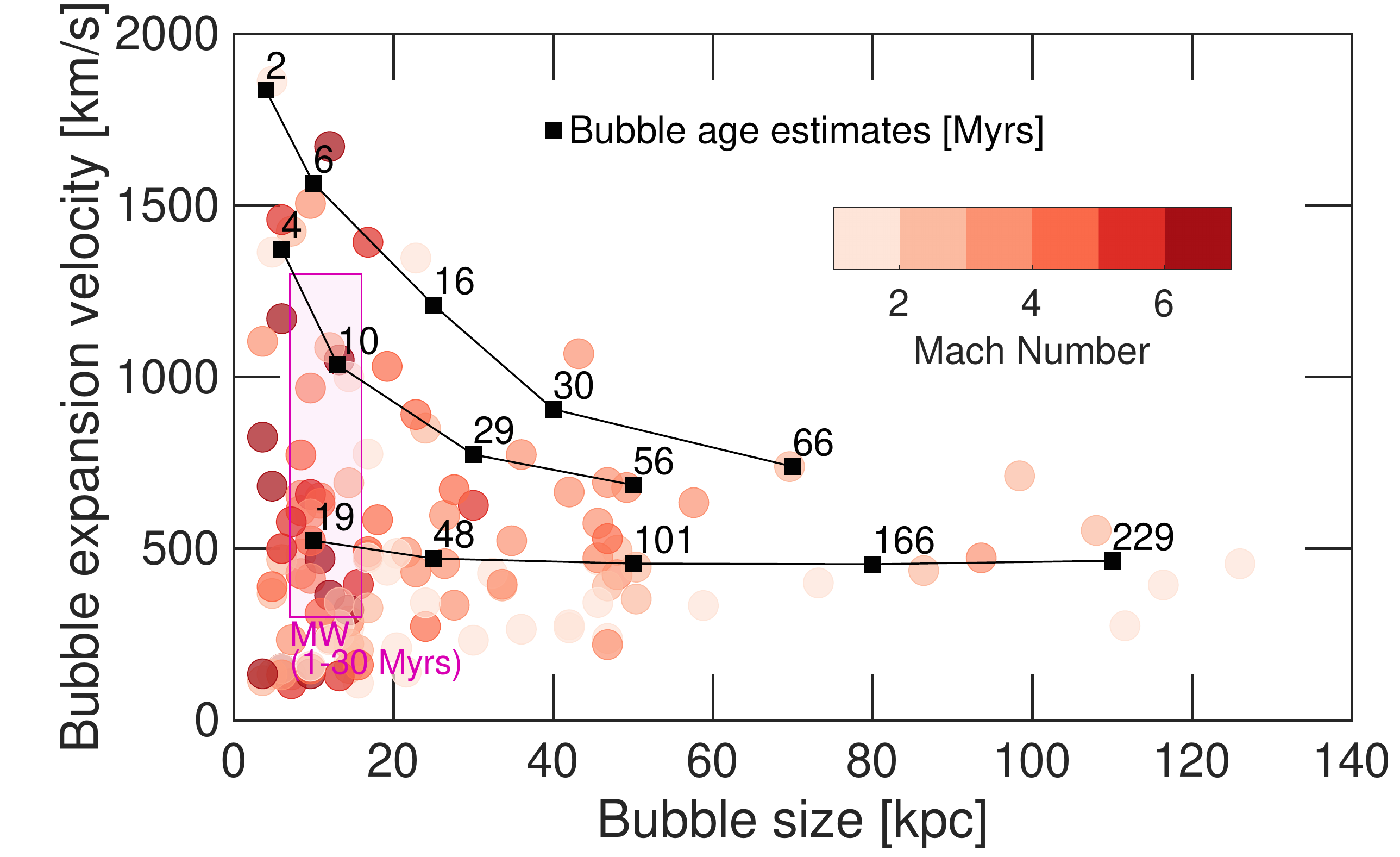}
    \caption{Radial outflow velocities of the gas and bubble expansion speeds in the TNG50 MW/M31-like galaxies with CGM bubbles at $z=0$. Top left: profiles of the radial outflow velocities of the gas as a function of galactocentric distance for the galaxy depicted in Fig.~\ref{fig:top2_2}, Subhalo ID 535410: curves denote the 50th, 90th, and 95th percentiles of the velocity distributions. Top right: distribution across the galaxy sample of the maximal (i.e. 95th percentile) of the gas outflow velocities at different distances. Main bottom panel: bubble expansion speed vs. bubble size, for the smallest bubbles of the 127 TNG50 galaxies with bubbles. Here we assume that the 99th percentile of the outflow velocities of the gas between the center and $1.2$ times the bubble size is a good proxy for the bubble expansion speed. Each galaxy is represented with one circle, color coded by the mass-weighted average shock Mach number recorded for gas within the same radial location. Black annotations indicate the estimated ages of the TNG50 bubbles, based on the average velocity and size of the bubbles in the plot region and on the sampled ``trajectories'' (black squares and lines, see text for details). 
    The magenta rectangle denotes observational estimates for the Galaxy, including literature estimates of their ages.
    In TNG50, gas within the bubbles and at the bubble fronts can flow outwards with radial velocities of up to $1000-2000~\KMS$, and typically with weakly supersonic shock fronts.}
    \label{fig:velocities}
\end{figure*}

Here we characterize in more detail the outflow velocities of the gas within and around the bubbles. It should be clear from the bottom left panels of Figs.~\ref{fig:top2_1} and \ref{fig:top2_2} that to summarize the complex and diverse velocity fields of any galaxy with just a few numbers is going to be a difficult, if not arbitrary, task: we warn the reader that many possible missteps may be lurking when comparing to observed cases. Furthermore, in the following we measure outflow speeds of all gas, without distinguishing between warm and hot phases.

An example galaxy is represented in the top, left panel of Fig.~\ref{fig:velocities}: it is the same galaxy of Fig.~\ref{fig:top2_2}. We measure in spherical geometry the instantaneous, mass-weighted, radial outflow velocity of the gas as a function of its galactocentric distance. Because within each galaxy, and at any radial shell, gas can move with a range of speeds, the profiles are given for the 50th, 90th, and 95th percentiles of the velocity distributions in each bin of distance: thick black, thinner gray, thin light gray curves, respectively. Furthermore, the profiles are shown separately for the gas that outflows above and below the galactic disk, as indicated. On average, the gas in this galaxy outflows with speeds of up to a few tens to a few hundreds $\KMS$, depending on distance. However, the tails at high speed of the distributions clearly show discontinuities, i.e. large sudden variations, with gas in relatively thin radial shells moving as fast as many hundreds or even $\gtrsim2000~\KMS$. The location of the high-velocity peaks roughly traces the location of the pressure edges in the maps of Fig.~\ref{fig:top2_2}, but not in all cases.

The top right panel of Fig.~\ref{fig:velocities} extends the quantification of the high-velocity outflows to the whole galaxy sample studied here: different histograms show the distribution across the TNG50 galaxies with bubbles of the 95th percentile of their gas outflow velocities, for gas at different distances: within 5 kpc, between 5 and 15 kpc, between 15 and 25 kpc, from black thick to thin light gray curves. Different galaxies show different maximal outflow velocities, but generally it can be seen that maximal outflows are somewhat slower at larger distances, at least within a few tens of kpc distance from the centers. For the median (16th-84th percentile) galaxy studied here, the 95th percentile of the gas in the innermost regions outflows at about $202~\KMS$ ($107-622~\KMS$); however, about a dozen galaxies in the sample are caught, at the time of inspection, as their fastest gas flows outwards faster than $1000~\KMS$. For reference, $200-400~\KMS$ (purple annotations) are the velocity estimates of a constant radially-outward wind model that fits the kinematic MaNGA data of the so-called Akira galaxy \citep{Cheung.2016}: whereas this is redder than the galaxies in our sample, it is the poster child of the 'red-geysers', whose large-scale AGN-driven winds are akin to those in IllustrisTNG. We also include the estimates of the outflow velocities for the Makani galaxy \citep{Rupke.2019}: 200 and 1500 $\KMS$ out to about 50 kpc above and below Makani's disk, respectively (red annotations). Whereas the Makani galaxy is thought to be a starburst induced by a galaxy merger, and so it is not in fact a good match to the sample we are focusing on here as it is also more massive, we include it as a reference point because it is the only example we are aware of for outflows that produce cocoon-like morphologies similar to the MW bubbles, albeit in cool ionized gas. Within our sample, we find no correlation between the highest tails of the gas outflow velocities and the global properties of the galaxies, such as their stellar or total halo mass, or SMBH mass. This could possibly be due to the fact that we focus on a relatively narrow range of galaxy masses and/or that the bubbles are de facto stochastic phenomena whose physical properties and states at a given time do not need to correlate with galaxy properties, like SMBH mass, that vary on much longer timescales.

The velocity profile in the top left panel of Fig.~\ref{fig:velocities}, together with analogue ones we have inspected, suggests that the highest velocity gas is often confined in relatively thin radial shells of a few kpc thickness, particularly at small galactocentric distances. Even if in general gas outflow velocities and bubble expansion (i.e. shock) speeds are {\it not} the same, we deduce that the high tails of the gas outflow velocity distributions could be taken as a reasonable proxy for the speed at which the bubbles expand. We have partially verified this assumption by directly measuring the outflow velocity distributions of the gas located in radial shells of some kpc around the recorded size of the previously-identified bubble features (where the sizes are the ones discussed in Section~\ref{sec:sizes}). The correspondence between the peak heights of the $v_{\rm out,95}$ or $v_{\rm out,99}$ radial profiles and the instantaneous maximal radial outflow velocity of the gas around the bubble front holds to within $10-100~\KMS$ in many cases, albeit not all: we use this correspondence further in Fig.~\ref{fig:velocities}.

In the main lower panel of Fig.~\ref{fig:velocities}, we show estimates of the bubble expansion speeds in TNG50 against the physical extent of the bubbles: one dot per galaxy, one bubble (the smallest) per galaxy. In the absence of finely-spaced simulation outputs for all interesting galaxies and in the absence of an automated bubble identification method, such estimates are obtained from the 99th percentiles of the mass-weighted, instantaneous outflow velocity distribution of gas located between the galactic center and $1.2$ times the size of the smallest bubble of each of the 127 TNG50 MW/M31-like galaxies that exhibit at least one dome-like feature. 
For bubbles smaller than $40-50$ kpc, the galaxy-to-galaxy variation is large: bubbles can expand radially as fast as $\sim2000~\KMS$, but also as slow as about $100~\KMS$. For bubbles smaller than 20 kpc, i.e. whose extent is more comparable to that of the eROSITA and Fermi bubbles in the Galaxy, the median TNG50 bubble at $z=0$ expands at about $500~\KMS$. The rather broad observational constraints on the expansion speed of the Galaxy's bubbles, or rather on the kinematics of the warm or cold gas clouds therein \citep[magenta rectangle, e.g. from ][]{Fox.2015, Bordoloi.2017, Ashley.2020, DiTeodoro.2018}, are well within the values predicted by TNG50. However, from the distribution of Fig.~\ref{fig:velocities} it is clear that larger bubbles expand typically at lower velocities than smaller bubbles or at least do not exhibit the very high velocities of some of the small bubbles. This does not imply that low-velocity bubbles go the farthest, but rather that bubbles do not expand with constant speed and in fact some of them seem to slow down as they expand, as we explicitly show later in Figs.~\ref{fig:timeevol_1} and \ref{fig:timeevol_2} -- there we follow individual outflow and bubble events as a function of time instead of taking a population snapshot.

We also color-code every TNG50 bubble with the median Mach numbers of the gas cells at the same radial location where the velocities are estimated and undergoing shocks. As already seen in Figs.~\ref{fig:top2_1}, \ref{fig:top2_2}, and \ref{fig:top30_machnum}, shock fronts can develop as a consequence of the pushing of the gas from the innermost regions of the galaxy and of its gaseous halo, even if the most coherent shock fronts in our images are those at tens of kpc distances. Here we quantify that those shock fronts are weakly supersonic, and the typical TNG50 bubble develops shocks with $1.8-3.7$ average Mach numbers (25th-75th percentiles). On the other hand, the smallest bubbles of each galaxy, which are typically the fastest and are depicted in Fig.~\ref{fig:velocities}, develop shocks with median Mach numbers of about $3.2$ but also as high as $7-10$.


\subsection{Estimates of the bubble ages}
Based on the distribution of the bubble expansion velocities vs. bubble size of the main panel of Fig.~\ref{fig:velocities}, we can estimate the age of the bubbles at the time of inspection: namely, based on what is available in the simulation output, we can estimate how long ago the individual bubbles started to inflate. 

Our bubble age estimates are annotated in black text in the main panel of Fig.~\ref{fig:velocities}, for representative loci (black squares) in the bubble expansion speed vs. size plane. As noted above, the collective behaviour of the TNG50 bubbles at $z=0$, i.e. the distribution of points in Fig.~\ref{fig:velocities}, supports the idea that gaseous bubbles do not expand with constant speed, but rather slow down as they get larger, i.e. as the gas moves further out in the halo. This is at least the case for the {\it smallest} bubble in each halo at any given time and for the fastest bubbles of that ensemble\footnote{We have noticed that the maximal outflow velocities of the gas around or within the {\it largest} bubbles of multiple systems are on average a factor of two higher than those of the smallest bubbles, at fixed bubble size and for sizes $\gtrsim40-50$ kpc: we speculate that, as consecutive energy injections take place and bubbles pile up in the halo, the largest ones are also pushed from the gas that is in turn pushed by more recent energy injections. However, a dedicated study is required to properly delineate the dynamics and phenomenology of galaxies with systems of multiple and successive bubbles and shells.}. We hence assume the expansion speed of the bubbles to vary with time (i.e. size) along the black `trajectories' in the main panel of Fig.~\ref{fig:velocities}: these tracks represent the fastest (95th), fast (84th), and average bubbles (50th percentiles) of the TNG50 velocity distributions in bins of bubble sizes, from top to bottom. In fact, the average bubbles exhibit an almost constant-velocity expansion.

According to TNG50, bubbles of about 10 kpc in size or height and that currently move at $1500~\KMS$ ($1000~\KMS$) can be about 6 (10) million years old. These estimates are again in the ball park of the disparate values put forward for the Milky Way, where for an inferred size of about 10 kpc and inferred expansion velocities varying from about 300 to 1300 $\KMS$, the age of the Galaxy's bubbles has been estimated to be of about some million years and overall in the range between 1 and 30 Myrs (see Introduction). Based on the TNG50 results and under the assumption that the same mechanism produces bubbles both in TNG50 and in the Galaxy, the very young and very old extremes put forward for the Galaxy appear less frequent or plausible and the intermediate observationally-derived age estimates (of e.g. $6-9$ Myr by \citealt{Bordoloi.2017}, but also up to 20 Myr) more in line with our modeling.

For larger bubbles, e.g. $40-60$ kpc sized bubbles, ages can vary from about 30 to about 100 million years. However, it should be noted that, a priori, a 50 kpc bubble with instantaneous expansion speed of a few hundred $\KMS$ may have expanded in its past along faster (or slower) tracks than the ones we are sampling with the black lines and squares in Fig.~\ref{fig:velocities}.

Given the simulation data at hand, it is not possible to reconstruct the past evolution of all the TNG50 galaxies studied thus far with a time cadence better than about 150 Myr (see Section \ref{sec:sample} for details). 
Therefore, the immediate past history of most of the bubble features discovered so far at the $z=0$ snapshot remains elusive. However, we can study the time evolution with a time spacing of just about a few million years for a few selected galaxies: this is what we are going to do in the next Section.

\section{Time evolution and connection to SMBH activity}
\label{sec:smbhfeedback}

We have demonstrated so far that, in the cosmological simulation TNG50, over-pressurized, dome-like features of gas in the CGM above and below the disks of MW and M31-like galaxies are an emergent and rather frequent phenomenon of the underlying galaxy formation model. However, we have not yet addressed the nature of the energy injections from the galactic centers that give rise to such gaseous patterns. In the following, we demonstrate that, within the TNG50 model, bubbles whose appearance is similar to those seen in X- and $\gamma$-rays in the Galaxy are produced by kinetic, wind-like energy injections driven by the SMBHs at the galaxy centers.

\begin{figure*}
		\includegraphics[width=17.8cm]{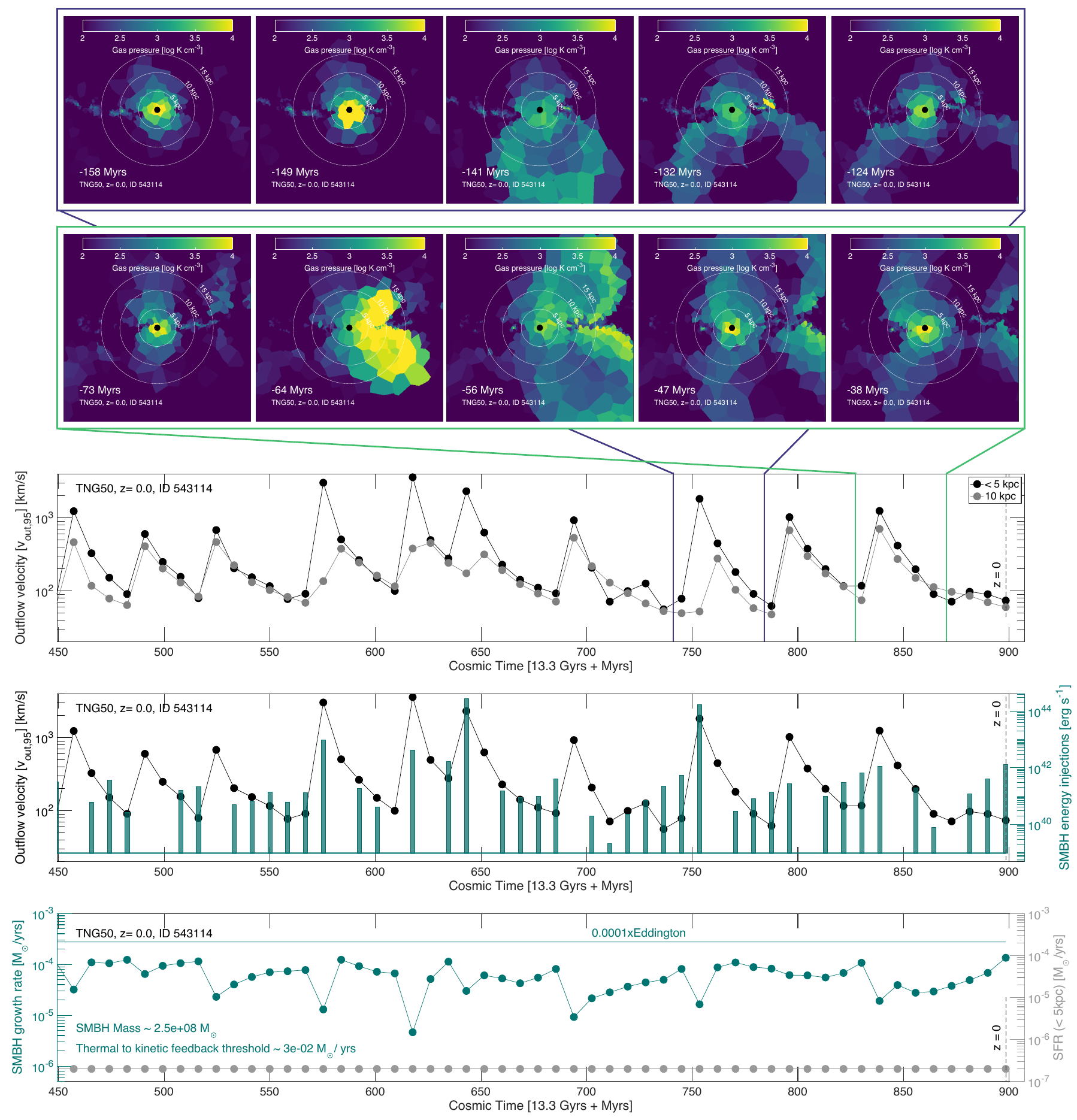}
    \caption{Time evolution of selected quantities pertaining to the TNG50 galaxy with Subhalo ID 543114 at $z=0$. The top two rows show the evolution across 34 Myr of two bubble events, seen in edge-on projections: the maps depict the cross-section of the Voronoi tessellation along a plane perpendicular to the galactic disk, passing through the center, and color-coded by the mass-weighted gas pressure. The lower three panels quantify the time evolution of, from top to bottom, outflow velocities for gas at different galactocentric distances; one of the latter as it compares to the time evolution of the energy injected by the SMBH between two successive snapshots, i.e. across time spans of 8.5 Myrs; the instantaneous mass accretion rate of the SMBH and the SFR within 5 kpc apertures. In all cases, the depicted time interval spans about 450 million years in the past of the selected galaxy: the current epoch at $z=0$ is at the rightmost end of each panel.}
    \label{fig:timeevol_1}
\end{figure*}

\begin{figure*}
		\includegraphics[width=17.8cm]{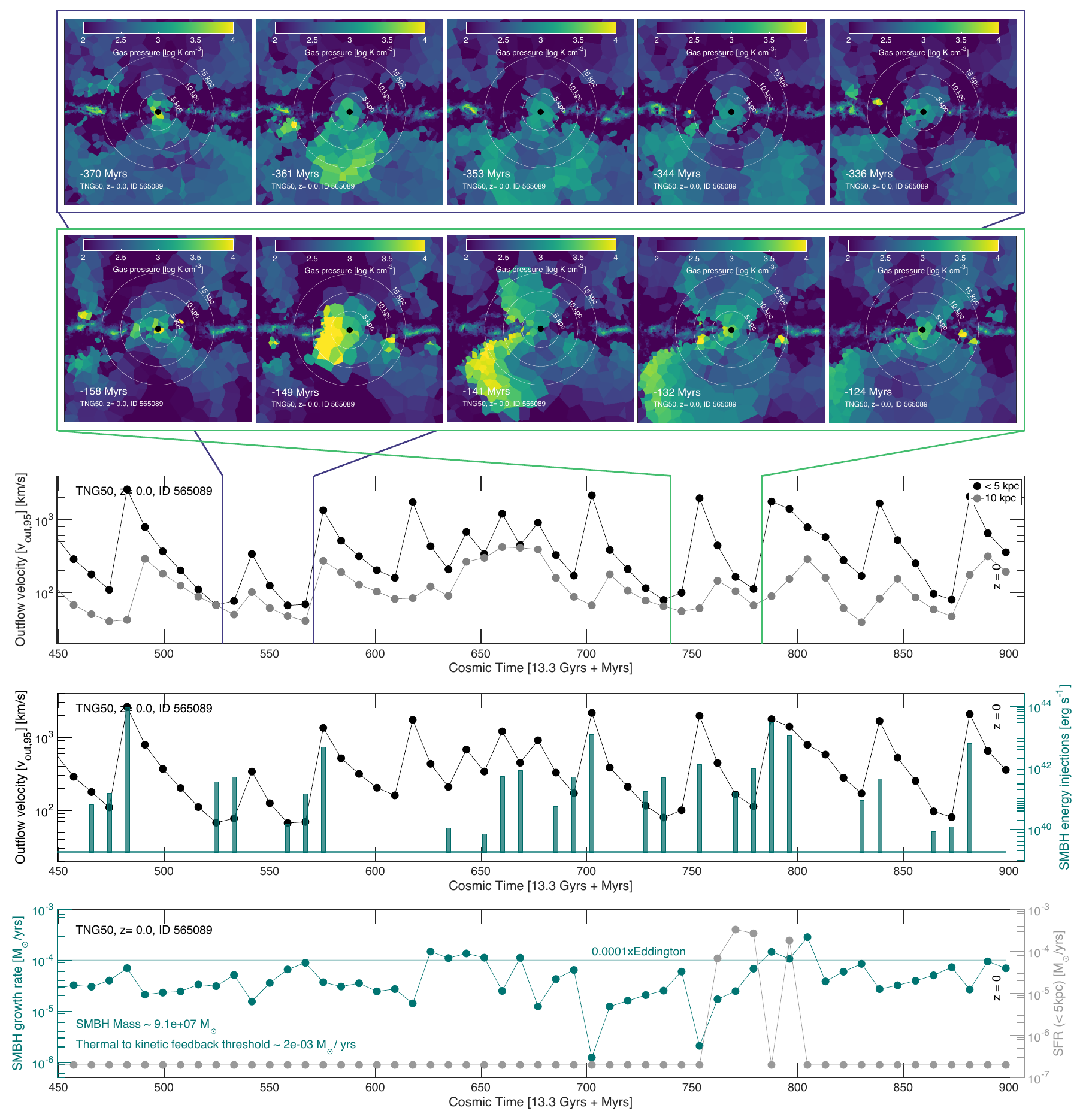}
    \caption{As in Fig.~\ref{fig:timeevol_1} but for the TNG50 galaxy with Subhalo ID 565089, which at $z=0$ exhibits four over-pressurized bubble-like features above and below the galactic disk. Both from here and the previous Figure, we can see a clear time correlation among the development of over-pressurized bubbles, the emergence of high-velocity outflows and the injections of energy from the SMBH at the center.}
    \label{fig:timeevol_2}
\end{figure*}

\subsection{The recent past of two TNG50 example galaxies}
Thanks to the data stored in the so-called subboxes of TNG50 (see Section~\ref{sec:sample} and \citealt{Nelson.2019release}), we can follow the recent past history of a couple of simulated galaxies in great detail. In particular, two galaxies in the TNG50 MW/M31-like sample that host bubble-like features in their circumgalactic gas at $z=0$ are located in the so-called Subbox2 and can thus be studied at a time cadence of 8.5 Myr and for $\sim1$ Gyrs prior to $z=0$.

The time evolution of these two example galaxies (Subhalo IDs 543114 and 565089) is shown in Figs.~\ref{fig:timeevol_1} and \ref{fig:timeevol_2}, respectively. 

In the top rows of Figs.~\ref{fig:timeevol_1} and \ref{fig:timeevol_2}, for each galaxy, we show two bubble events, across five snapshots each (i.e. across a time span of about 34 Myrs): the mass-weighted gas pressure is shown from the edge-on projections of the galaxy, the black filled circle at the center representing the location of the galaxy's SMBH. The maps actually show the Voronoi tessellation of the simulated galaxies, through a cross-section perpendicular to the galactic disks and passing through the centers. 

The three plots in the lower panels of each Figure, on the other hand, quantify the time evolution of selected quantities, whereby the x-axis denotes cosmic time in Myr starting from 13.3 Gyr after the Big Bang. In practice, the depicted time spans about 450 million years in the past of the selected galaxies, the current epoch at $z=0$ being at the rightmost end of each panel.
From top to bottom, we show the time evolution of: a) the outflow velocities of the gas at different galactocentric distances; b) the outflow velocities of the gas in the innermost regions as their evolution compares to the energy injected by the SMBH at the center of each galaxy between two subsequent snapshots; and c) the instantaneous growth rate of the SMBH and the instantaneous SFR measured from the gas within the central 5 kpc, both in solar masses per yr. 

\subsubsection{Episodic bursts of outflowing gas}

The time evolution of the outflow velocities of the gas clearly shows that, in TNG50, the gas is ejected  in an episodic manner. The gas in the innermost 5 kpc (black curves and circles) is accelerated to radial outflow velocities of up to a few thousands $\KMS$ every $20-50$ million years. After peaking at such outflow speeds, it then progressively exhibits lower velocities at subsequent times. The gas at larger distances (between 5 and 10 kpc, gray curves and circles) shows a very similar time modulation, however with lower velocity peaks and, in some instances, with visibly delayed maxima. A similar time series for the gas at even larger distance (omitted to avoid overcrowding the plot) would show a consistent trend: progressively lower and more delayed outflow velocity peaks and maxima. The time evolution of the outflow velocities of Figs.~\ref{fig:timeevol_1} and \ref{fig:timeevol_2} corroborates what we have deduced from the population study of Fig.~\ref{fig:velocities}: namely, the outflows slow down towards larger galactocentric distances, at least within a few tens of kpc from the galaxy centers. This is consistent with the general findings of \cite{Nelson.2019} for TNG50 galaxies across the mass spectrum and redshift.

The sequence of images in the upper portions of Figs.~\ref{fig:timeevol_1} and \ref{fig:timeevol_2} shows how individual bubbles develop and how they correspond to the outflow velocity peaks of the evolution plots.  Over-pressurized features of gas develop quickly between subsequent snapshots (i.e. in less than 8.5 Myrs) and expand in one or the other direction. For example, for the bubble event depicted at the top of Fig.~\ref{fig:timeevol_2}, the pressure patterns show a bubble whose edge reaches a radius of about 20 kpc in $\lesssim 8.5$ Myrs, implying an expansion velocity of about 2-3 kpc Myr$^{-1}$, i.e. about $2000~\KMS$: this falls at the high-end of the distribution of the expansion speeds for small bubbles presented in Fig.~\ref{fig:velocities} for the $z=0$ TNG50 galaxy sample. In the second sequence of images of Fig.~\ref{fig:timeevol_1}, on the other hand, we can see the development of a bubble that appears to expand both above and below the galactic disk: in this case, the orientation of the energy injection must have been somewhat parallel to the disk plane. Importantly, this image suggests that the same bubble feature seen from a different edge-on projection may actually appear as a vertically-oriented pair of bubbles (as e.g. those in the second sequence from the top of Fig.~\ref{fig:timeevol_2}), albeit not necessarily symmetric.

\subsubsection{Suppressed star formation in the inner regions of typical TNG50 galaxies}
\label{sec:nosfr}
In TNG50, the energy that pushes the gas is sourced by the SMBHs at the galaxy centers. In fact, even admitting that the non-local stellar feedback implemented in TNG50 could be the cause for the gas phenomenology quantified so far -- which we think not possible by construction --, the SFRs in the centers of the MW/M31-like galaxies are typically very low, if not vanishing. 

For the high-cadence galaxy data, the instantaneous SFR of the gas within the innermost 5 kpc is shown in the bottom panels of Figs.~\ref{fig:timeevol_1} and \ref{fig:timeevol_2}, gray curves and circles: there is no time modulation. In fact, the SFR in the inner regions of the two galaxies is, at most times in their recent past, vanishing and placed by hand at the lowest limit of the depicted range. We have checked also the SF histories from the ages of the stellar particles that are in the innermost regions at $z=0$ (which form at a temporal resolution as good as $\sim0.03$ Myrs, or worse): for these two galaxies, the stellar particle-based SF histories are too consistent with essentially no star formation for the depicted span of time. While there is disk gas within 5 kpc or smaller radii, this is not star forming because it is below the density threshold for star formation in our model: we think the gas is diluted by the same action of the SMBHs that drive the large-scale outflows. 
Therefore, feedback from star formation is not the mechanism for the formation of the TNG50 bubbles depicted in Figs.~\ref{fig:timeevol_1} and \ref{fig:timeevol_2}. 

We have checked that negligible levels of SF are the case not only for the galaxies of Figs.~\ref{fig:timeevol_1} and \ref{fig:timeevol_2}, but more generally for the $z=0$ sample of 127 MW/M31-like galaxies that feature bubbles. We have measured the instantaneous SFRs and specific SFRs of the gas within 1, 2, and 5 kpc apertures around the galaxy centers at $z=0$ and find (albeit do not show) that in more than $70-90$ per cent of the studied galaxies -- depending on the aperture -- the star formation in their central regions is lower than 0.01 $\MSUN$ yr$^{-1}$, i.e. negligible. While there is typically some gas in the simulated galaxies within these inner regions, it is not star forming because it is not dense enough. We have checked this to be the case also for SFRs averaged over a few to a few tens of Myr and by inspecting the SF histories of the stellar particles at the centers over the past $\sim500$ million years: the majority of the TNG50 MW/M31-like galaxies with bubbles have suppressed star-formation in their center at recent epochs. However, there are a number of TNG50 galaxies with bubbles (about one or two dozen, depending on the considered aperture) where the instantaneous SFR in the inner regions is larger than 0.1 $\MSUN$ yr$^{-1}$. The corresponding sSFRs are consistent with a starburst, with sSFR$\sim10^{-9}$yr$^{-1}$. These TNG50 systems are interesting cases that potentially offer similarities with the Galaxy, where the recent ($\lesssim 30$ Myrs) SFR in the central molecular zone or within a few tens to a few hundreds of pc from the center has been estimated to be in the range $0.1-0.8~\MSUN$ yr$^{-1}$ \citep{Yusef-Zadeh.2009, Nogueras-Lara.2020}. We have also checked the recent SF histories of the stars in their central regions and find a handful of cases with ``bursty'' SF in the last few tens of Myr within the inner $\sim500$ pc, i.e. with histories possibly more similar to that of the Galaxy. However, the relatively low frequency of TNG50 galaxies with non-negligible SF at their centers implies that feedback from star formation is generally not the mechanism responsible for inflating the bubbles in TNG50.

\subsubsection{Bubbles inflated by separate and subsequent events of energy injections from the SMBHs}
\label{sec:smbh_origin}

The connection between over-pressurized bubbles, high outflow velocities and feedback from the SMBHs is shown in the middle time-series plots of Figs.~\ref{fig:timeevol_1} and \ref{fig:timeevol_2}. There the time evolution of the 95th percentile of the outflow velocities in the innermost region of each galaxy (black curves and circles) is contrasted to the energy injected by their SMBHs between two subsequent snapshots, i.e. across a time span of 8.5 Myr (teal columns)\footnote{As we have not recorded the timing and energy of the individual SMBH feedback injections in the simulation output, here we estimate both quantities by measuring the total (internal and kinetic) energy gained by the gas cells within a small fixed distance from the center (here 3 kpc) since the previous snapshot, i.e. over the previous 8.5 Myrs: this implies that the mechanical luminosities of Figs.~\ref{fig:timeevol_1} and \ref{fig:timeevol_2} (teal columns in the second panels from the bottom) are only approximate, by a factor of a few to one order of magnitude.}.

The energy released from the SMBHs into the gas and the peaks of high-velocity outflows are well-synchronized, strongly pointing to causality. Strictly speaking, in our model, there would be no other possible source of energy in the galaxy but for the feedback from the SMBHs. In the case of the two galaxies of Figs.~\ref{fig:timeevol_1} and \ref{fig:timeevol_2}, the SMBH feedback is in the form of mechanical feedback at low-accretion rates, whereby the gas surrounding the SMBH is given kicks in random and different directions at different energy injections: we show this explicitly in the bottom panels of Figs.~\ref{fig:timeevol_1} and \ref{fig:timeevol_2}. 

The masses of the SMBHs of these two galaxies is $2.5\times10^8~\MSUN$ and $9\times10^7~\MSUN$, respectively, and their growth is negligible over the half billion year of the inspected time evolution. The instantaneous growth rate of the SMBHs sampled at time intervals of about 8.5 Myr is shown as a teal curve and circles in the bottom panels Figs.~\ref{fig:timeevol_1} and \ref{fig:timeevol_2}: it reads $10^{-4}-10^{-6}~\MSUN$yr$^{-1}$. The dips in the SMBH growth rates that appear often synchronized with the outflow velocity peaks (upper panels) are a manifestation of the self-regulated nature of the SMBH feedback and growth model in IllustrisTNG, as the accretion into the SMBHs is hampered (at least temporarily) if gas is vacated from the innermost regions of the simulated galaxies. For reference, in the same panels, we show the corresponding Eddington mass accretion rate for the two galaxies (essentially a constant) multiplied by a factor of $10^{-4}$: these two galaxies are firmly in a very low-accretion rate state. In fact, the SMBHs of these two galaxies accrete mass at a rate that is one or two orders of magnitude lower than the threshold, given their SMBH mass, that in our model determines whether feedback is in thermal or kinetic mode (see Section~\ref{sec:model} and Equation~\ref{eq:chi}): the two galaxies of Figs.~\ref{fig:timeevol_1} and \ref{fig:timeevol_2} have not exercised thermal mode feedback since at least 1 Gyr in the past (we have checked past the depicted time span).

The mechanical feedback luminosities that are released as pulsated injections in our simulated galaxies vary in the $10^{40-44}$ erg s$^{-1}$ range, whereby $\epsilon_{\rm f, kin}\times6\times 10^{42}$ erg s$^{-1}$ is the feedback energy accumulated for a SMBH accreting at a rate of $10^{-4} ~\MSUN$ yr$^{-1}$ and where the coupling efficiency $\epsilon_{\rm f, kin}$ is 0.2 at most in the kinetic regime in our model \citep[see  Section~\ref{sec:model} and ][for more details]{Weinberger.2017}. These values, and those anticipated in Section~\ref{sec:model}, are in the ballpark of the wind or mechanical powers invoked by more idealized but more sophisticated simulations of AGN-driven (Fermi) bubbles \citep[e.g.][]{Yang.2012, Guo.2012, Mou.2014}. 

On the other hand, and differently than what is normally assumed in analytical and numerical models so far, based on the analysis of the two galaxies in Figs.~\ref{fig:timeevol_1} and \ref{fig:timeevol_2}, in TNG50 it would appear that, typically, a few consecutive but separate energy releases or activity bursts build up together to inflate individual (pairs of) bubbles. We believe that the ensuing periodicity of the outflows, and hence of the bubbles, is due to a {\it combination} of modeling choices and physical conditions within the simulated galaxies, as we argue below. 

As mentioned in Section~\ref{sec:model}, there are two modeling choices (and associated parameters) that determine, respectively, the amount of energy converted and stored into kinetic feedback and the times and hence amounts of the pulsated energy depositions. These are 1) the aforementioned coupling efficiency $\epsilon_{\rm f, kin}$, which is capped at $0.2$ but which can be much smaller depending on the density of the gas immediately surrounding the SMBHs, with associated density-threshold parameter (as per Equation 9 of \citealt{Weinberger.2017}); and 2) the burstiness parameter of Equation 13 of \citealt{Weinberger.2017}, which determines the frequency of the reorientation of the kinetic kicks, depending on the gas mass enclosed  in the feedback region and the depth of the potential well. A larger value for this parameter would imply that fewer feedback events occur, but each individually stronger. The motivations for these choices and for the fiducial values of the corresponding parameters are given in the already-cited method paper.

However, it is important to note here that, whereas these choices do influence the outcome in the simulated galaxies, they do not directly determine it in any obvious manner, as the manifestations of the model tightly depend on the physical conditions of the gas and of the SMBHs within the galaxies, which in turn depend on their past history, both immediate and remote. This should be clear by noticing two facts (based on the teal columns in the middle evolutionary plots of Figs.~\ref{fig:timeevol_1} and \ref{fig:timeevol_2}): within the same galaxy, the amounts of energy deposited in the pulsated kinetic kicks vary by up to $3-4$ orders of magnitude across injection events; across galaxies (or at least between the two we can inspect), the timings between consecutive energy injections (or at least the ones we can capture) vary between 8 and 50 Myr -- this with the same underlying modeling choices. Finally, whether strong, high-velocity outflows burst out of the galaxy centers (black solid curves and dots in the top two evolutionary plots of Figs.~\ref{fig:timeevol_1} and \ref{fig:timeevol_2}), in turn producing the over-pressurized CGM bubbles, also seems to vary, as multiple energy releases appear to be required to trigger high-velocity outflows: these manifest themselves at time intervals in the range $20-50$ million years.

\subsection{The long-term fate of the bubbles}
\label{sec:bubblesend}
Interestingly, the $z=0$ pressure, X-ray, temperature and Mach number maps spanning 200 kpc a side of the galaxy of Fig.~\ref{fig:timeevol_1} (not shown) clearly depict the vestiges and patterns associated with three bubble events, most certainly the three most recent episodes, dating back to about $150-160$ million years ago -- see e.g. the outflow peaks in Fig.~\ref{fig:timeevol_1}. Similarly, the $z=0$ pressure, X-ray, temperature and Mach number maps spanning 200 kpc a side of the galaxy of Fig.~\ref{fig:timeevol_2} (not shown) reveal four bubble-like patterns, also in this case corresponding to the four most recent events of the last $150-160$ million years of evolution. TNG50 enables us to conclude that bubble-like outflow features are in fact most common, and most expected, around roughly Milky Way-mass galaxies with disk-like morphologies; moreover, according to TNG50, other features of piled-up gas in coherent fronts may be present in the CGM of the Galaxy, at larger galactocentric distances than those probed by the eROSITA and Fermi bubbles. However, based on the visual inspection of the galaxies in Fig.~\ref{fig:timeevol_1} and ~\ref{fig:timeevol_2}, it would appear that the typical ``lifetime'' of individual bubbles is $150-200$ million years at most, at least for features searched and recognizable within galactocentric distances of up to about 100 kpc above and below the galactic disks. We have confirmed this by visually inspecting the past history of all the galaxies in our sample at a time cadence of about 150 Myrs: firstly, the same features are not recognizable across such time spans; secondly, as it can also be appreciated from the maps throughout this paper, quantities like the X-ray luminosity, pressure, and temperature in the dome-like shells appear to decrease for larger bubbles until a point where no feature can be identified. We postpone to future work the tasks of physically characterizing the long-term fate of the bubbles. But it may be fortunate that the Fermi/eROSITA bubbles have not yet propagated to a distance where they may be too dim to be observed (see Discussion Section~\ref{sec:discussion}).

\begin{figure}
		\includegraphics[width=8.5cm]{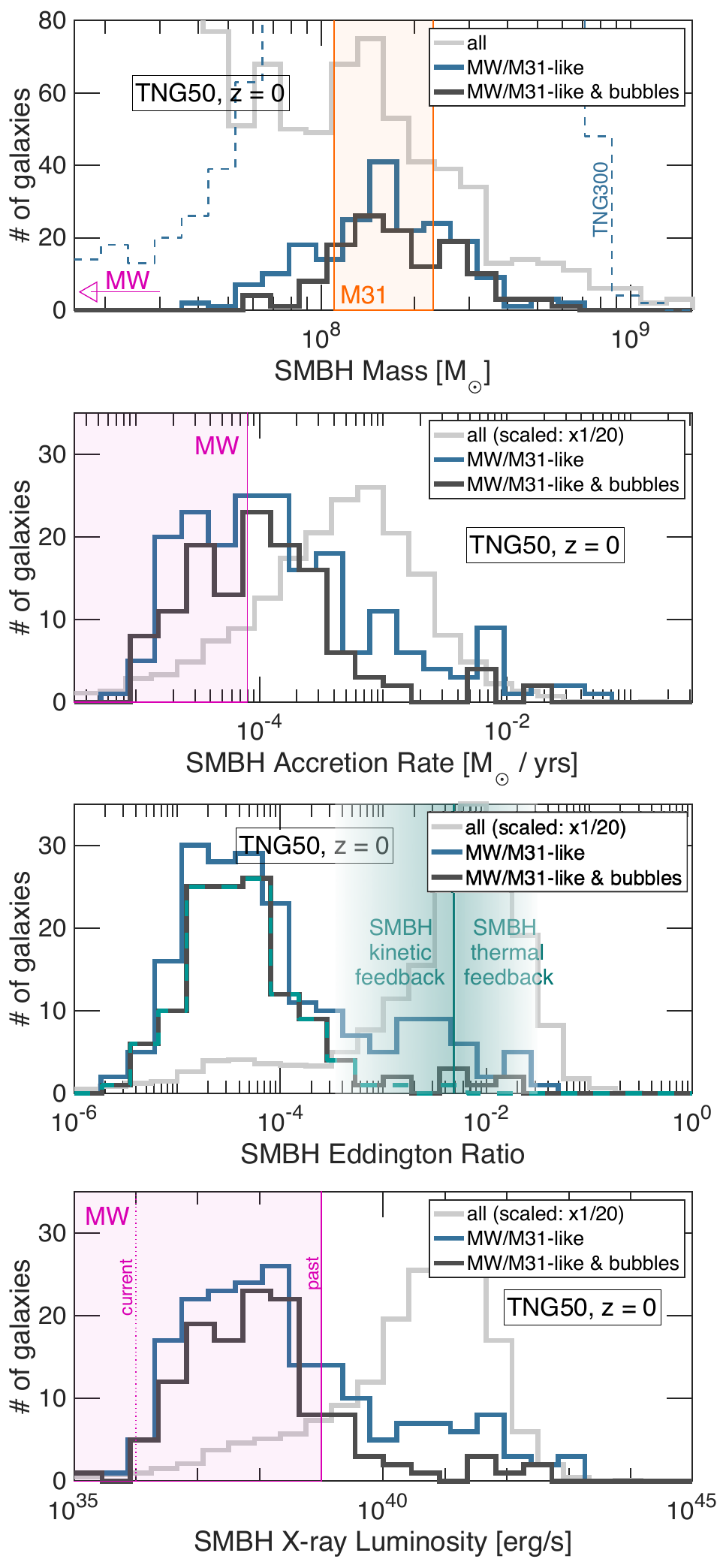}
    \caption{Demographics of SMBHs in TNG50 at $z=0$. From top to bottom, we show the distributions of the mass, mass accretion rate, Eddington ratio, and X-ray luminosity of the SMBHs in all TNG50 galaxies (light gray), TNG50 MW/M31-like galaxies (as selected in Section~\ref{sec:sample} -- blue), and TNG50 MW/M31-like galaxies that exhibit dome-like features above and below their stellar disks (as identified in Section~\ref{sec:pressure} -- dark gray histograms). Observational constraints on the Galaxy's and Andromeda properties are in magenta and orange annotations (see text for details). In the top panel, the dashed blue histogram shows the distribution of the SMBH masses in the TNG300 simulation. In the third panel, the dashed teal histogram shows the distribution of the Eddington ratios of the TNG50 MW/M31-like galaxies with bubbles that are in kinetic feedback mode. The great majority of MW/M31-like galaxies with bubbles host SMBHs that exercise kinetic i.e. SMBH-driven wind feedback, and accrete at low Eddington ratios such that they are, therefore, not necessarily highly luminous.}
    \label{fig:smbhs}
\end{figure}

\subsection{SMBHs and their activity in TNG50 galaxies with bubbles}
\label{sec:smbh_demographics}

The picture presented above exemplifies the phenomenology of the TNG50 bubbles: episodic energy injections from the SMBHs at the center of MW/M31-like TNG50 galaxies push the gas out of the innermost regions of the galaxies and of their CGM, developing high-velocity outflows and often producing weakly supersonic shock fronts well into the far reaches of the halo. This is yet another manifestation of SMBH feedback in action very similar to the one we have described in idealized setups in \cite{Weinberger.2017}, their Fig. 1, and to the one we have uncovered in higher-redshift IllustrisTNG galaxies in \cite{Nelson.2019}, their Figs. 2 and 3. 

It is important to point out that such energy injections are sourced by weakly-accreting SMBHs, and therefore not highly luminous ones. The mass, accretion rate, Eddington ratio, and X-ray luminosity distributions of the SMBHs in the TNG50 galaxies at $z=0$ are quantified in Fig.~\ref{fig:smbhs}, where MW/M31-like galaxies (blue histograms) and MW/M31-like galaxies exhibiting bubble features (dark gray histograms) are compared. In the top panel, the dashed blue histogram shows how a larger sample of MW/M31-like galaxies from the TNG300 simulation includes lower mass SMBHs than the smaller TNG50 volume, i.e. in the tails of the distribution. The mass accretion rates and the Eddington ratios are instantaneous, i.e. they reflect the state of the SMBHs at the time of inspection, even if they may have inflated their last bubbles many tens, if not hundreds, of million years before. However, we have checked and the corresponding histograms of $10^{10.5-11.2}~\MSUN$ galaxies in TNG50 at a few snapshots prior to $z=0$, i.e. across about half a billion years in the past, are essentially indistinguishable from to those of Fig.~\ref{fig:smbhs}. In the bottom panel, the X-ray luminosity of the simulated SMBHs is derived by following \citealt{Churazov.2005} to obtain the bolometric luminosity, i.e. by distinguishing between radiatively efficient and inefficient AGNs, and by adopting the correction factors proposed by \cite{Hopkins.2007} to convert from bolometric to hard+soft X-ray luminosity -- the details are given in e.g. Section 4.1.1 of \citealt{Habouzit.2019}. However, it is important to keep in mind that how the mass accretion rates of SMBHs, which are the physical quantities predicted by the simulation, convert into luminosities is highly uncertain and is subject of extensive study for both observations and theoretical applications.

As it can be seen from the middle panels of Fig.~\ref{fig:smbhs}, the majority of the MW/M31 analogues in TNG50 have their SMBHs growing at low-accretion rates and, as such in our model, injecting kinetic rather than thermal feedback. The bulk of the TNG50 MW/M31-like galaxies, and of those that exhibit bubbles, host SMBHs that accrete at $\lesssim 10^{-3}~\MSUN$ yr$^{-1}$, namely with Eddington ratios lower than $0.1$ per cent and with X-ray luminosities lower than $10^{39-40}$ erg s$^{-1}$. In particular, in TNG50, bubble features are identified less frequently in galaxies whose SMBHs are in the thermal-feedback mode, i.e. at relatively higher Eddington ratios: whereas 83 per cent of the MW/M31 analogues in TNG50 are in kinetic feedback mode at $z=0$, this fraction raises to 95 per cent for the MW/M31-like galaxies with bubbles. Now, within our sample of MW/M31-like galaxies, SMBHs at high-accretion rates are typically found in galaxies of lower mass, or lower SMBH mass: as we have seen that bubble features are less frequent among lower-mass galaxies in our sample (top right panel of Fig.~\ref{fig:demographics}), they are also less frequent in galaxies with smaller SMBHs (top panel of Fig.~\ref{fig:smbhs}). 

In the panels of Fig.~\ref{fig:smbhs}, magenta and orange annotations denote observational inferences about the SMBH mass of the Galaxy and Andromeda and estimates of Sgr A* accretion rate and X-ray luminosity. It is clear from the top panels of Fig.~\ref{fig:smbhs} that in TNG50 we have no MW/M31-like galaxy whose SMBH is as small as that of the Galaxy, whose mass is about $4\times10^6\MSUN$ and hence more than one order of magnitude smaller than that of the SMBH in Andromeda (see Section~\ref{sec:discussion} for a discussion). However, a large fraction of TNG50 galaxies with bubbles host SMBHs that accrete at low rates similarly as to the Galaxy's, at least according to the estimates by \citealt{Quataert.1999}, who place upper limits at $\lesssim8\times10^{-5}\MSUN$ yr$^{-1}$, indicated in the second panel from the top of Fig.~\ref{fig:smbhs}. However, other estimates place the {\it current} mass accretion rate of Sgr A* at $10^{-9}-10^{-7}\MSUN$ yr$^{-1}$ \citep[e.g.][]{Agol.2000, Marrone.2006}. In fact, how the current luminosity of Sgr A* e.g. estimated via modeling from e.g. polarised infrared and 
X-ray flares \citep{Yuan.2004} translates into the average accretion rate over its past $\sim$ tens of Myr remains uncertain and model dependent \citep[e.g.][]{Genzel.2010}. Moreover, should Sgr A*  be fueled entirely by stellar mass loss \citep{Genzel.2010}, then the comparisons in Fig.~\ref{fig:smbhs} would not be meaningful. However, the upper estimates for the current, steady, bolometric luminosity of Sgr A* are $\lesssim 10^{37}$ erg s$^{-1}$ \citep{Narayan.1998}: these would correspond to $\lesssim 10^{36}$ erg s$^{-1}$ in X-ray (soft + hard bands), assuming the latter contributes by about 10 per cent. Other measurements are as low as $10^{33}$ erg s$^{-1}$ in X-ray, i.e. $10^{-11}$ times the Eddington luminosity \citep{Baganoff.2003}. Yet, arguments have been proposed that the mass accretion rate and the X-ray luminosity of our Galaxy's SMBH have been higher in the past, by even $3-4$ orders of magnitude \citep{Totani.2006}. For example, it has been suggested via X-ray reflection nebulae that Sgr A* might have been brighter a few hundreds years ago, with a luminosity of a few $10^{39}$ erg s$^{-1}$ \citep{Revnivtsev.2004, Ponti.2010}. Within the framework and limitations of our modeling, we include estimates of both the current and past properties of Sgr A* in the bottom panel of Fig.~\ref{fig:smbhs}. For radiatively inefficient AGNs, a fraction of TNG50 SMBHs may shine as dimly as the Galaxy's.

Amid the complex framework outlined above, our model shows that features like the ones seen in the Galaxy could be in principle sourced by non-highly energetic SMBHs (in fact with accretion rates as low as $10^{-6}- 10^{-3}$ times the Eddington limit, third panel from the top of Fig.~\ref{fig:smbhs}), so long as the ejected energy couples effectively to the surrounding medium. In conclusion, according to our results, bubble-like features from low-accretion state AGNs are easily produced, as demonstrated by a model for SMBH kinetic feedback even as crude as the one in TNG50. Figs.~\ref{fig:top2_1} and \ref{fig:top2_2} show that the emerging morphology can even be quite close to that of the bubbles in the Galaxy, albeit this does not seem to be always the case.

\section{Discussion, interpretations, and implications}
\label{sec:discussion}

\subsection{Connection to our Galaxy, and on SMBH feedback as physical origin of its bubbles}
\label{sec:disc:MW}
The qualitative and large-scale morphological similarities between the CGM bubbles of some of the simulated galaxies in TNG50 and those that have been seen in the Galaxy in X-ray, $\gamma$-rays, in microwaves, and in polarized radio emission (see Introduction) are certainly evocative and encouraging, and grant further investigations. Further, the resemblance uncovered in this paper offers an indirect argument for favoring an AGN-related origin of the Galaxy's bubbles. 

In fact, the over-pressurized, dome-like features of gas in the CGM above and below the disks of MW and M31-like galaxies in TNG50 are an {\it emergent} phenomenon of feedback from SMBHs, one for which the underlying IllustrisTNG galaxy formation model has not been in any way calibrated nor previously tested for. Even though the TNG50 simulation has been designed to model galaxies in general, and not to model our Galaxy specifically, in this paper we have found that, not only can we identify galaxies hosting bubbles with total galaxy stellar mass and star formation rate similar to those of the Galaxy's (Fig.~\ref{fig:demographics}), but also cases of bubbles that extend for sizes (Fig.~\ref{fig:sizes}) and expand with speeds (Fig.~\ref{fig:velocities}) that are in the ball park of estimates for our Galaxy. We have also seen that, although within TNG50 the overwhelming majority of MW analogues with bubbles are sourced by the activity of the SMBHs, with negligible SF in their centers (Sections~\ref{sec:nosfr}), there are examples of TNG50 MW-like galaxies with bubbles that also exhibit active and complex SF histories within the innermost regions resolved by our numerical scheme, which are good candidates for similarities with the case of our Galaxy.

With this paper we are therefore laying the groundwork for possible future analyses, by putting forward TNG50 as a rich laboratory to study the emergence of the Galaxy's Fermi and eROSITA bubbles under the boundary conditions of a specific, but realistic, galaxy formation model and in the context of our finding that these are a manifestation of feedback from SMBHs, specifically SMBH-driven winds. 
Furthemore, TNG50 provides dozens of MW analog galaxies, as well as galaxies at different evolutionary stages, by hence offering plausible and diverse test beds for working hypotheses and interpretations of observational facts, and for the verification of results derived from more idealized setups and calculations.
For example, the width of the $\gamma$-ray and X-ray emission at  the galactic disk level, i.e. at the base of the Galaxy's bubbles, has been qualitatively used as a way to discriminate among SF winds, AGN-driven winds and AGN jet scenarios as the physical cause for the Fermi bubbles \citep{Zhang.2020}. Here we notice that broad bases of the X-ray emission at the galactic disk level can easily be obtained via AGN-driven winds: namely, TNG50 simulated galaxies seen from an external view point naturally return also many kpc-wide bases of the X-ray emission, even if the bubbles are inflated by SMBH activity. Whereas this may be thought as due to the still limited resolution of the TNG50 subgrid feedback implementation and whereas further work is required to make claims on the $\gamma$-ray emissions, the bases of the X-ray emission maps are diverse across simulated galaxies, and typically proportional to the bubble size. 

However, amid the connections to our Galaxy highlighted thus far, it is clear, e.g. from the details of Section~\ref{sec:smbh_demographics}, that there is much more knowledge about our own Galaxy than what the TNG50 simulation can address. In particular, the finite numerical resolution and the limitations of the adopted effective modeling (Section~\ref{sec:methods}) do not allow us to capture what occurs in the central few tens of parsecs of galaxies and, even less so, in our own Galactic Center. The latter is known to be a site of great activity, with rapid episodes of stellar disk formation in the last few Myr \citep{Genzel.2006} and with Sgr A* being thought to be fed by accretion of gas \citep{Cuadra.2006} from  winds of the young massive stars in the inner $\sim0.5$ pc \citep{Paumard.2006}. This phenomenology is unattained within the framework of TNG50. However, the results of this paper still can provide useful insights also for the case of our own Galaxy. In particular, it has been generally argued that AGN cannot be the cause of the eROSITA and Fermi bubbles -- assuming they have been sourced by the same physical phenomenon -- because of the current low-luminosity of Sgr A* \citep[][and previous Section]{Genzel.2010}, and hence of the possibly low inferred level of energy available for feedback, at least at present. However, firstly, as mentioned above, within the picture of radiatively inefficient accretion flows, a number of lines of evidence are consistent with the average accretion rate of Sgr A* over the last tens of Myr being three or four orders of magnitudes higher than its current one \citep[see a compilation by][]{Mou.2014}; secondly, within the TNG50 framework we have shown that SMBHs at low-accretion rates (as low as $10^{-6}-10^{-3}$ Eddington ratios) can drive high-velocity outflows, and therefore bubbles, the actual situation in the Universe depending on how exactly the energy ejected from the innermost regions of galaxies couples to the surrounding medium.

\subsection{The case for larger bubbles, and of M31 and other galaxies}

The results presented in the previous Sections support a picture whereby, firstly, if the mechanism that inflated the bubbles in the Galaxy is episodic as is the case in TNG50, other features of piled up gas in coherent fronts may be present in the CGM of the Galaxy, at larger galactocentric distances than those probed by the eROSITA and Fermi bubbles.
Secondly, X-ray bubbles similar to, or more extended than, the ones seen in the Galaxy could be a rather common feature of disk-like galaxies. If the bubbles in X-ray and $\gamma$-ray are sourced by the same dynamical origin, e.g. energy injections sourced by the SMBH, also $\gamma$-ray bubbles similar or more extended than the Fermi bubbles could in principle be present in external disk-like galaxies. 

In particular, within TNG50 MW/M31 analogues, coherent, dome-like features of over-pressurized gas that impinge into the CGM of galaxies are apparently less frequent in starbursts and fully-quenched galaxies, and so more frequent in main-sequence disk-like galaxies or those in the green valley (see Section~\ref{sec:demographics}). As we postpone to future efforts the task of developing an automated identification method of bubbles and of extending the analysis to simulated galaxies at higher redshift, across a wider mass range, and across multiple projections, we conjecture that there are two fundamental requirements for coherent bubble features to manifest themselves in the CGM, in gas pressure, temperature or X-ray luminosity: the presence of a sufficiently-dense gaseous disk capable of re-directing, in a bipolar-like fashion, the outflows that are sourced from the innermost regions of galaxies; and the presence of a sufficiently-dense, stratified gaseous halo against which such outflows can hit, possibly develop shocks, and dissipate (as in the idealized tests of \citealt{Weinberger.2017}). We speculate that galaxies like the Milky Way and Andromeda, i.e. disk-like, star-forming or green-valley galaxies in $10^{12}~\MSUN$ haloes, constitute the sweet spot for such phenomena to manifest themselves, whereas e.g. in higher-mass or already fully-quenched galaxies, energy injections from SMBHs may produce randomly-oriented ``jets'' of material. 

Now, evidence for extended $\gamma$-ray emission from the halo of Andromeda has been discussed and confirmed in the literature \citep{Pshirkov.2016, Ackermann.2017}: however, whether this can be associated with bubble pairs, i.e. quasi-spherical features symmetrically located perpendicular to the M31 galactic disk remains debated \citep{Karwin.2019}. On the X-ray emission side, it has been argued that a substantial mass of the CGM in M31 should be present in its hot gaseous diffuse atmosphere \citep{Lehner.2015, Bregman.2018}: however, whether X-ray features in the halo of Andromeda as the ones predicted here could be detected with future telescopes and observations remains to be assessed. Certainly, our findings motivate further efforts, both theoretical and observational, in the direction of our neighboring M31, and more generally in the local Universe -- as we outline below. 

\begin{figure}
		\includegraphics[width=7.5cm]{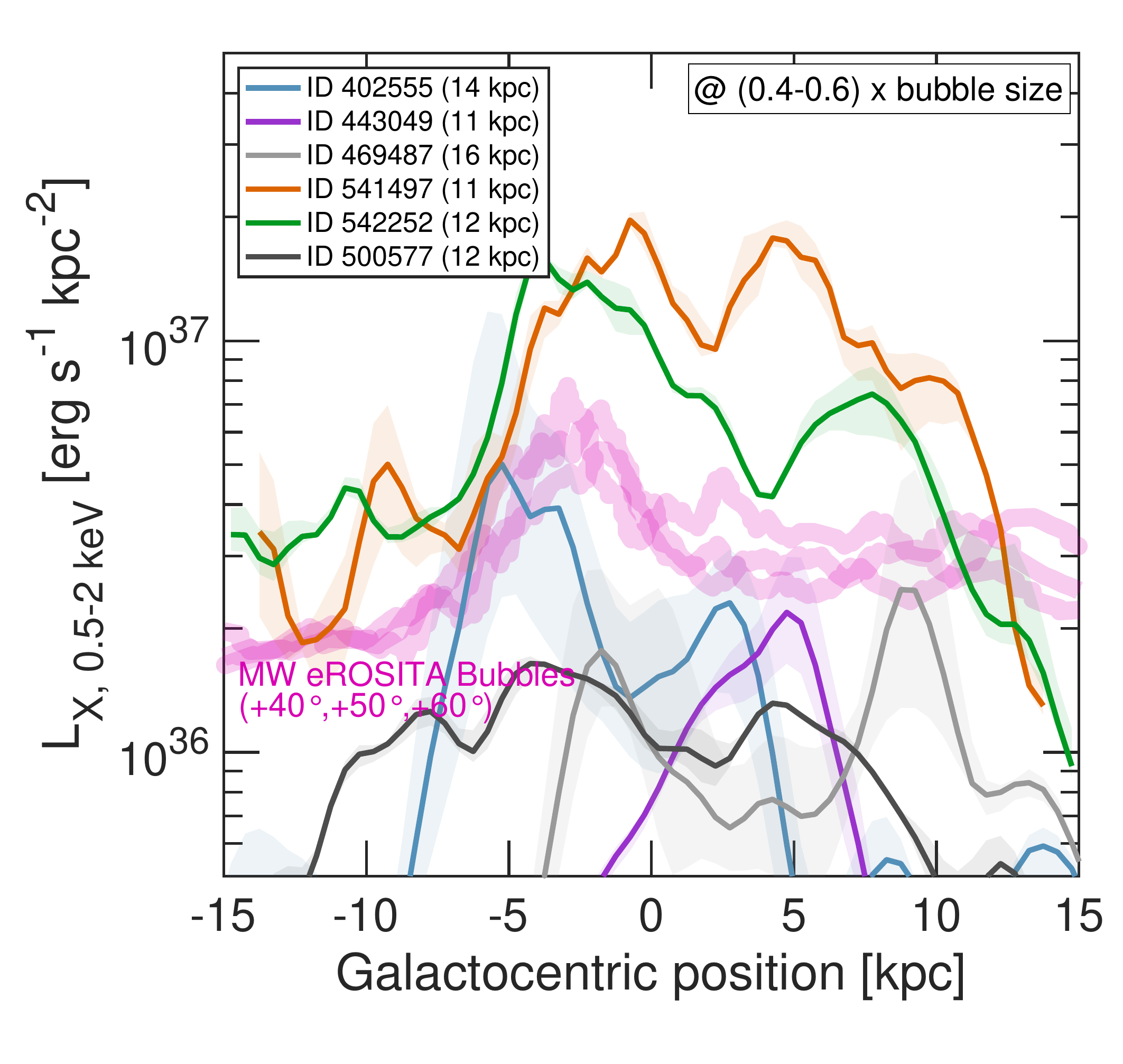}
		\includegraphics[width=7.5cm]{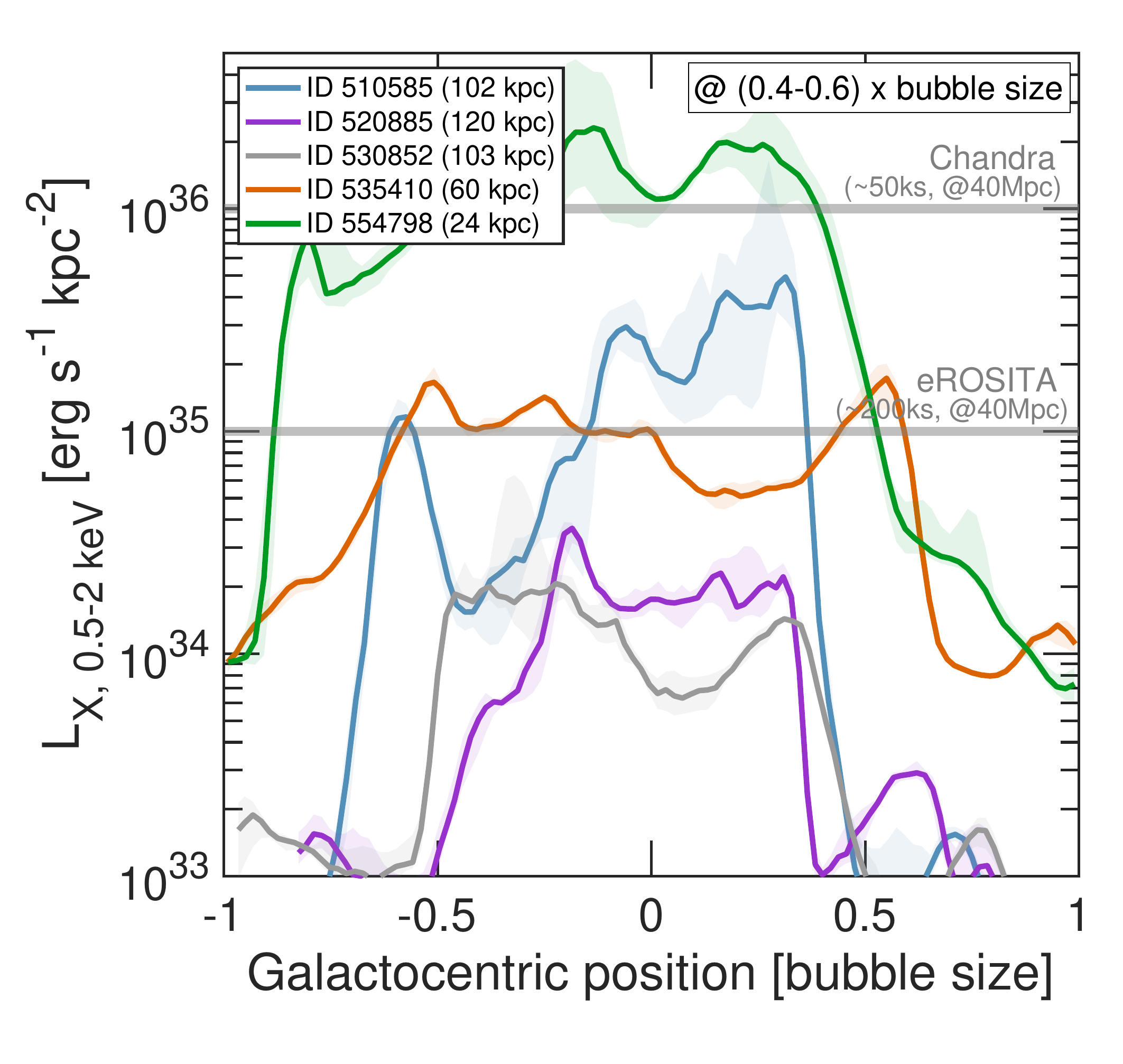}       
    \caption{One-dimensional intrinsic X-ray surface-brightness profiles of a selection of TNG50 galaxies with bubbles in edge-on projections, at the fixed height of about half the bubble size and as a function of projected radius from the galaxies' centers. In the top panel, we show the X-ray modulation of a few TNG50 galaxies with bubble sizes similar to that of the eROSITA bubbles (see Fig.~\ref{fig:sizes}). The shaded magenta profiles are, for mere reference, those of the Galaxy, as measured by eROSITA in the Northern hemisphere \citep{Predehl.2020}, arbitrarily normalized. In the bottom panel, we select a few TNG50 galaxies whose stellar mass and SFR are compatible with the Galaxy's (see Fig.~\ref{fig:demographics}) and measure their X-ray profiles across their largest bubble. The sizes of the bubbles are given in the legends, in kpc. Gray lines indicate the current and expected surface brightness limits of observations with Chandra and eROSITA, respectively, for galaxies placed at e.g. 40 Mpc distance (see text for details). From X-ray emission, often, but not always, the TNG50 bubbles exhibit geometrical morphologies that are consistent with shells or cavities of diverse thicknesses and diverse elongations in the projected plane.}
    \label{fig:x-ray_profiles}
\end{figure}

\subsection{Preliminary X-ray properties and detectability}
\label{sec:x-rayprops}
We cannot claim that TNG50 yields a fully realistic model for the Galaxy's bubbles (see discussion of Section~\ref{sec:disc:MW}); however, here we believe it interesting and useful to expand upon the qualitative similarities mentioned in the previous Sections in relation to the X-ray properties of the gas above and below the galactic planes. 

In Fig.~\ref{fig:x-ray_profiles}, we give the intrinsic X-ray luminosity profiles of selected TNG50 galaxies with bubbles and show to what degree the X-ray morphologies of Fig.~\ref{fig:top30_Xray} may be consistent with shells or cavities. In particular, we measure the X-ray emission from the gas at a constant height of about half the bubbles' height ($\pm$ 10 percentage points), as a function of galactocentric position in one given edge-on projection of a few TNG50 MW/M31-like galaxies. Here we do not account for observational realism, in that e.g. we neglect line-of-sight absorption and we do not mock any particular observational setup. 

In the upper panel of Fig.~\ref{fig:x-ray_profiles}, we show the profiles of a few TNG50 galaxies whose bubble size is similar to that of the eROSITA bubbles (see Fig.~\ref{fig:sizes} and Section~\ref{sec:sizes}): we have chosen to show in Fig.~\ref{fig:x-ray_profiles} examples whose X-ray angular modulation is reminiscent of that observed in the Galaxy, e.g. at intermediate and high latitudes: the magenta shaded bands in Fig.~\ref{fig:x-ray_profiles} are the high-latitude X-ray profiles from Fig. 2 of \cite{Predehl.2020}. As we do not account for geometric and projection effects nor mimic the eROSITA observations, these are arbitrarily rescaled and are not meant for direct quantitative comparison but only for qualitative reference. Yet, in this plot, for the TNG50 galaxies we account only for the contribution of gas within 20-kpc deep layers, de facto excising the contribution of the whole gaseous haloes. The profiles in the top panel of Fig.~\ref{fig:x-ray_profiles} show that bubbles whose X-ray morphology is consistent with that of projected (quasi-)spherical shells, and with the Galaxy's, do exist in TNG50. However, the thickness of the shells -- i.e. the width of the X-ray luminosity enhancements -- can be very diverse. The magnitude of the modulation, which necessarily also depends on the height at which the one-dimensional profiles are measured, varies from galaxy to galaxy, between less than a factor of two to a factor of a few. 
We also notice that the same bubble may show somewhat different X-ray one-dimensional profiles depending on the exact projection. 

In the lower panel of Fig.~\ref{fig:x-ray_profiles}, we show the same profiles but for a selection of TNG50 galaxies whose stellar mass and global SFR are consistent with those of the Galaxy, as described in Section~\ref{sec:demographics} and shown in Fig.~\ref{fig:demographics}. However, here we show the X-ray profiles of selected galaxies across their largest bubble, i.e. possibly at heights of about 100 kpc: for the comparison across different galaxy-bubble systems, we hence normalize the projected galactocentric positions in units of the considered bubble size. To connect more directly to observations of external galaxies, here we account for all the gas that is gravitationally bound to the galaxies, and therefore all gas along the line of sight: the central troughs representing the inner cavities would be more enhanced if we had shown the signal from only a thin layer of circumgalactic gas, as done in the top panel. As for the examples in the top panel, there are TNG50 bubbles whose X-ray morphology is consistent with under-luminous cavities surrounded by higher-luminosity layers. The lower panel of Fig.~\ref{fig:x-ray_profiles} shows that this can be the case also in the far reaches of the gaseous haloes, that the magnitude of the X-ray angular modulations in individual systems can be as large as many factors also in the full projected signal, and that the shapes of shells can be diverse, from more spherical ones (e.g. Subhalo ID 535410, with higher-luminosity layers at $\pm 0.5$ the bubble size, also in Fig.~\ref{fig:top2_2}) to more vertically- elongated ones (e.g. Subhalo ID 520885).

In Fig.~\ref{fig:x-ray_profiles}, lower panel, shaded gray bands denote estimates of the surface brightness limits of X-ray observations of galaxies in the local Universe (e.g. within 40 Mpc distance) with the Chandra telescope for about $50$-ks exposures ($10^{36}$ erg s$^{-1}$ kpc$^{-2}$) and with eROSITA for $200$ ks (down to $10^{35}$ erg s$^{-1}$ kpc$^{-2}$). These estimates are obtained from existing Chandra observations of nearby MW/M31-like galaxies \citep[e.g. from][]{LiWang.2013} and from eROSITA predictions for extended sources in the local Universe by \citealt{Merloni.2012} and \citealt{Oppenheimer.2020}, respectively. For comparison, eROSITA should be able to detect $10^{36}$ erg s$^{-1}$ kpc$^{-2}$ with about 25 ks of exposure time. As most of the TNG50 galaxies have X-ray surface brightness above $10^{35-36}$ erg s$^{-1}$ kpc$^{-2}$, the bubbles uncovered in this paper should be in principle well within the grasp of Chandra and eROSITA with reasonable observing time. We postpone to a dedicated paper the effort to quantify in how many cases, for what bubble sizes and for what X-ray halo luminosities shell-like features  could actually be identified in the X-ray maps of external galaxies, given the field of view and spatial resolution of available observatories and given the added complexity of projection effects.

\begin{figure}
		\includegraphics[width=8.5cm]{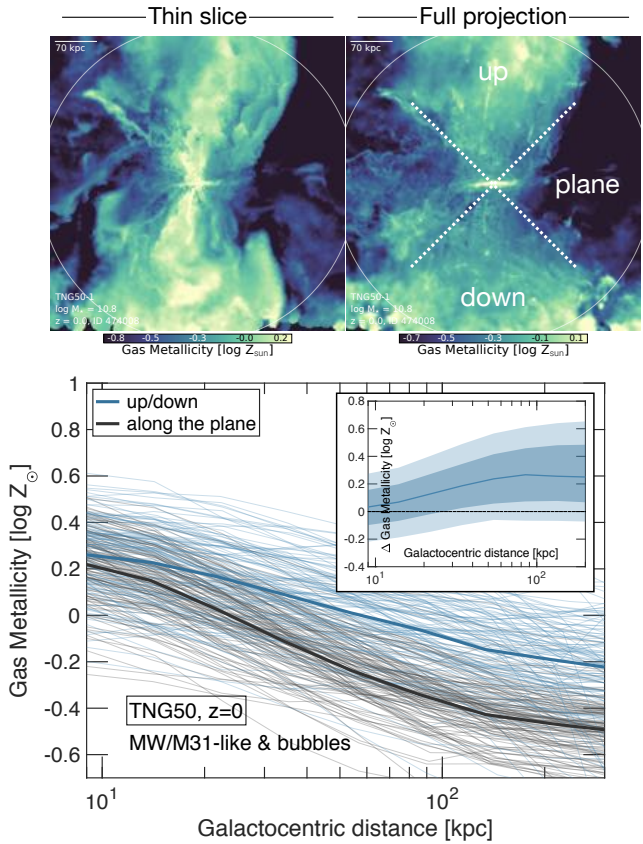}
    \caption{Angular dependence of the gas metallicity throughout the halo of MW/M31-like galaxies in TNG50 at $z=0$. The top panels show the mass-weighted gas metallicity in an example TNG50 galaxy with bubbles, seen edge-on, by contrasting a thin slice vs. the full projection across 500 kpc, the field of view of the maps. The main panel shows the 3D mass-weighted gas metallicity profiles for the gas within (beyond) 45 degrees from the galactic plane of the galaxies: along the plane (black) vs. up/down (blue) curves. Thin (thick) curves denote individual (median) profiles. The inset shows the logarithmic ratio of the profiles in the different directions from individual galaxies: on average (solid blue curve), the CGM above and below the disks of MW/M31-like galaxies is $0.2-0.3$ dex more enriched than along the galactic planes. Very similar average results are obtained for MW/M31-like galaxies with or without bubbles.}
    \label{fig:metals}
\end{figure}

\subsection{Other observational signatures of the TNG50 bubbles}

In other external, disk-like, edge-on galaxies, manifestations of SMBH feedback similar to those in TNG50 MW/M31-like galaxies could also be tested via X-ray stacking \citep{Truong.2021b}, by imaging the hardness of the diffuse X-ray emission \citep[as e.g. studied in the Sombrero galaxy][]{Li.2011}, or by probing the line ratios of highly-ionized species, such as OVII and OVIII and as done for the Galaxy \citep{Miller.2016}.

In fact, the angular directionality of the X-ray emission per se -- both the shell-like morphologies and the changes of X-ray brightness with galactocentric angle -- may ultimately be hard to detect in the nearby future when external or multiple galaxies are considered \citep[but see][]{Truong.2021b}: the same SMBH-driven winds that create the bubbles in MW/M31-like galaxies in TNG50 and overall the low-density, X-ray under-luminous regions preferentially above and below the disks of star-forming galaxies (e.g. Figs.~\ref{fig:top30_Xray}, \ref{fig:top2_1}, \ref{fig:top2_2}, \ref{fig:x-ray_profiles}) also eject highly-enriched material in columns of gas  that are perpendicular to the galactic planes: see e.g. bottom, second from the left, panels of Figs.~\ref{fig:top2_1} and \ref{fig:top2_2}. Therefore, probes of the gas metallicity in the CGM may be a more direct avenue to prove the ejective nature of SMBH feedback. 

We have already put forward theoretical predictions for the azimuthal dependence of the gas metallicity in  the CGM from the TNG50 simulation \citep{Peroux.2020}, but with a focus there on lower-mass galaxies and hence on stellar feedback. In Fig.~\ref{fig:metals}, we expand on those findings, and on the results presented thus far in this paper, to quantify the 3D mass-weighted gas metallicity profiles for MW/M31-like galaxies, out to about the virial radii of these galaxies, and in directions perpendicular and parallel to the simulated galactic disks. Firstly, the angular modulations of the gas metallicities of Figs.~\ref{fig:top2_1} and \ref{fig:top2_2} are in place also across halo scales e.g. across maps spanning 500 kpc per side: this is manifest from the example TNG50 galaxy in the top left panel of Fig.~\ref{fig:metals}. Secondly, the angular dependence is appreciable not only through thin slices, but also when the signal from the whole gaseous halo is averaged: left vs. right panel of Fig.~\ref{fig:metals}, top. Finally, the typical angular modulation of the gas in the CGM of MW/M31-like galaxies with bubbles is about $0.2-0.3$ dex, and maximal at about $\gtrsim 100$ kpc distances (see inset: solid curve, dark and light shaded areas denote the median, 16th-84th and 5th-95th percentiles of the distribution). Individual galaxies can exhibit CGM metallicity angular modulations as large as about 0.6 dex. These quantifications are consistent with the ones discussed in \cite{Peroux.2020}, but here we are certain that they are sourced by the ejective nature of SMBH, rather than stellar feedback. In fact, the presence itself of bubble-like features is not a prerequisite of the average trends of Fig.~\ref{fig:metals}, the only difference between galaxies with and without bubbles being a slight enhancement of the angular modulation at a few tens of kpc distances.

\subsection{On the modeling of SMBH feedback in TNG50}

Despite the unprecedented combination of numerical resolution and domain volume of the TNG50 simulation, the subgrid nature of the feedback mechanisms therein implemented at $\lesssim100$-pc spatial scales makes our modeling necessarily crude and simplified to various degrees. This is certainly the case for our modelling of the seeding, mass accretion and feedback injection of the SMBHs (see Section~\ref{sec:model}). However, here we argue that the details of how the AGN feedback is implemented are apparently not critical for the appearance of bubble-like features in MW/M31-like simulated galaxies, with two complementary lines of thoughts. 

Firstly, on the one hand and as already highlighted above, bubble-like features in the CGM of MW/M31-like TNG50 galaxies are yet another emergent manifestation of SMBH feedback in action, one that is very similar to the one we have uncovered in TNG50 in \cite{Nelson.2019}, their Figs. 2 and 3, but for low-redshift galaxies and mostly of late type. As already implied, that the large-scale outflows are preferentially aligned with the minor axis of the galaxies and that dome-like features develop above and below the galactic disk is likely simply a consequence of the outflowing gas following the path of least resistance imparted by the material in the inner, disky regions of galaxies and ploughing into the well-developed gaseous atmospheres. Thus, although the energy injection at small scales around the SMBHs does not have to follow any particular geometry, it naturally emerges in such a way on larger ($\gtrsim$ kpc) scales.

Secondly, the existence of features similar to ``bubbles'' is clearly not unique to the specific TNG50, i.e. IllustrisTNG, model for SMBH feedback. For example, in addition to the rich body of idealized experiments discussed in the Introduction and mentioned in this paper, recently, \cite{Costa.2020} have developed an implementation of small-scale AGN-driven winds in {\sc arepo} that is similar in essence to the low-accretion SMBH-driven winds in IllustrisTNG but that is implemented numerically in a more sophisticated fashion at the injection scale. Also such AGN-driven feedback produces gaseous morphologies that look suggestively similar to the eROSITA/Fermi bubbles (their Fig. 11) when applied to hydrodynamic simulations of idealized hydrostatic halos in Navarro-Frenk-White potentials of total mass $10^{12}~\MSUN$.

More generally, the fact that in the TNG50 modeling the thermal energy injections at high SMBH accretion rates do not develop high-velocity outflows, and possibly bubbles, is at least partially due to a limitation of our implementation: within our modeling, the energy injected at high-accretion rates and in the form of thermal energy is either radiated away very efficiently or impacts mostly gas that is star forming and, as such in our modeling, governed by an effective equation of state that sets an artificial temperature regardless of the amount of injected thermal energy \citep{Weinberger.2017, Zinger.2020}. We cannot exclude a priori that thermal-like and quasar-like energy injections may produce high-velocity outflows and bubbles. Nevertheless, the outcome we have unraveled from the TNG50 simulation in this paper shows that features like the ones seen in the Galaxy could be in principle sourced by non-highly energetic SMBHs, so long as the ejected energy couples effectively to the surrounding medium. 

As mentioned in Section~\ref{sec:smbh_demographics}, whereas the estimated mass of the SMBH in M31 is well represented by the TNG50 SMBH population, in TNG50 we have no MW/M31-like galaxy whose SMBH is as small as that of the Galaxy. 
Within the IllustrisTNG model and in general terms, the fact that in TNG50 no SMBH is so small for $10^{10.5-11.2}~\MSUN$ galaxies is not in itself a glaring failure of the model. It is in fact most certainly a combination of a) a possibly too high or too tight SMBH mass vs. galaxy mass relation emerging from our fiducial modeling \citep[as argued by e.g.][]{Terrazas.2020, Li.2020, Habouzit.2021} and b) a limitation of the sample statistics. The latter argument is supported by the dashed blue histogram in the top panel of Fig.~\ref{fig:smbhs}, which shows the SMBH mass distribution of MW/M31-like galaxies in the TNG300 simulation: in TNG300, with the same underlying galaxy formation model, lower resolution, but 200 times larger volume, many MW/M31 analogues have SMBHs as small as a few $10^6~\MSUN$ -- we count about 25 in the $3-6\times10^6~\MSUN$ range in TNG300 at $z=0$. Nevertheless, this issue is possibly a side effect of the IllustrisTNG model ``calibration'' and formulation, whereby prescriptions for SMBH mass growth, SMBH seeding mass and SMBH -- as well as stellar -- feedback are all interconnected and whereby some of the model parameters are degenerate. It is in principle possible to develop a model where the energetics of the SMBH feedback at $z=0$ remain the same and the (past) growth of the SMBHs return overall somewhat lower SMBH masses for MW/M31-like galaxies at $z=0$. 

In fact, it is the energetics of the feedback injections that is of relevance for the bubble phenomenology: we believe that similar outcomes as the ones presented in Figs.~\ref{fig:timeevol_1} and \ref{fig:timeevol_2} would be expected if the SMBH masses were $10-100$ smaller but their accretion rates were $10-100$ times larger. Therefore, if our model allowed for SMBHs as small as the one in the Galaxy and these were allowed to be in the low-accretion state, then the results in the example of Figs.~\ref{fig:timeevol_1} and \ref{fig:timeevol_2} would mimic what is happening in our Milky Way. In fact, within the IllustrisTNG model and e.g. in TNG300 even if not in  TNG50, the conditions above are met, albeit rarely (see e.g. Fig. 1 of \citealt{Zinger.2020}). Nevertheless, more sophisticated seeding and accretion implementations are certainly a key future modeling direction to alleviate some of the limitations of the IllustrisTNG model and to make it more likely to have MW analogues with lower SMBH masses than in TNG50.

Finally, higher-resolution simulations, dedicated numerical tests, or more tailored analyses would be needed to determine whether similar structures would be produced with continuous, rather than pulsed, SMBH-driven winds, and to certainly distinguish between contact discontinuities and shock features in the CGM of our simulated galaxies. In particular, additional dedicated simulations are needed to determine how predictive the time scales and frequency of the bubbles uncovered and modeled in TNG50 are (see Section~\ref{sec:smbh_origin}), specifically to quantitatively pin down the influence of our model parameters on the time scales of SMBH energy injection and hence on the frequency of the bubble events. Additional more focused analyses would also be needed to quantify how the energy content of the bubbles in TNG50 (including magnetic energy density in addition to thermal and kinetic gas energy) compares to the halo gas binding energies, how the bubbles are causally connected, in detail, to the quenching of star formation in the galaxies, and to quantify what the role of magnetic pressure and magnetic fields in general is in inflating the bubbles, if at all. These interesting directions -- beyond the scope of this first analysis of the TNG50 bubble phenomena -- would further deepen our understanding of the bubbles and their emergence in our multi-faceted galaxy formation model, and possibly in the Universe.

\section{Summary and outlook}
\label{sec:conclusions}

In this paper we have demonstrated that bubbles, shells, and cavities in the circumgalactic gas above and below the disks of Milky Way and Andromeda like galaxies -- and whose morphological features resemble those one seen in X- and $\gamma$-rays in our Milky Way -- are a natural outcome of the TNG50 cosmological simulation. 

In particular, we have focused on 198 MW/M31-like galaxies at $z=0$ that have been simulated within TNG50 and selected, among the thousands realized in the (52 Mpc)$^3$ volume domain, to have disky stellar morphology and a galaxy stellar mass in the $10^{10.5-11.2}~\MSUN$ range, and to be in relative isolation (Section~\ref{sec:sample}). We have visually inspected their gaseous content and properties from edge-on maps 200 kpc across at $z=0$ and find that about two thirds of such TNG50 galaxies exhibit one or more large-scale, coherent, dome-like features of over-pressurized gas that impinge into their gaseous haloes (Section~\ref{sec:pressure}). 
Our quantitative findings can be summarized as follows:

\begin{itemize}

\item The bubble features in TNG50 are prominent not only in gas pressure (Fig.~\ref{fig:top30_P_gas}), but also X-ray emission (Fig.~\ref{fig:top30_Xray}) and gas temperature (Fig.~\ref{fig:top30_T}), and often exhibit sharp coherent boundaries indicative of shock fronts (Fig.~\ref{fig:top30_machnum}).\\

\item The majority of TNG50 MW/M31-like galaxies that exhibit CGM bubbles include more than one feature, often in pairs above and below the galactic disks and more or less symmetric (e.g. Figs.~\ref{fig:voronoi_maps}, \ref{fig:top2_1} and \ref{fig:top2_2}). Some of the galaxies include a succession of bubbles or shells of increasing size, and overall these range from a few kpc to tens of kpc in height (Fig.~\ref{fig:sizes}). \\

\item Among the inspected MW/M31-like galaxies, TNG50 galaxies with bubbles are relatively less frequent at the low-mass end of that distribution ($\lesssim10^{10.6}\MSUN$ in stars); they are also less common in starbursts and in `fully' quenched galaxies than in main-sequence and green-valley systems (Fig.~\ref{fig:demographics}).
However, TNG50 galaxies with bubbles are relatively less frequent if they host SMBHs with mass accretion rates larger than $\sim 0.1$ per cent of the Eddington ratio for their mass (Fig.~\ref{fig:smbhs}).\\

\item The gas in the bubbles exhibits complex velocity and gas-metallicity fields (e.g. Figs.~\ref{fig:top2_1} and \ref{fig:top2_2}), with two fundamental properties. Firstly, the gas flows outwards mostly in directions perpendicular to the stellar and gaseous disks and with maximum ($95$th percentiles) radial velocities within $20-30$ kpc from the galactic centers of $100-1500~\KMS$ (10th-90th percentiles across the galaxy population, Fig.~\ref{fig:velocities}), but with peaks as high as $2000-3000~\KMS$ (Figs.~\ref{fig:timeevol_1} and \ref{fig:timeevol_2}). Secondly, the patterns of enriched gas follow those of the outflows, with prominent columns of $0.5-2~\ZSUN$ metallicity gas also directed perpendicularly to the galactic disks (Fig.~\ref{fig:metals}). \\

\item The bubble gas is hot, with mass-weighted temperatures of $10^{6.4-7.2}~$K (Fig.~\ref{fig:top30_T}), about one order of magnitude higher than the expected virial temperature for the $\sim10^{12}~\MSUN$ hosting haloes. The corresponding X-ray emission geometry (Fig.~\ref{fig:top30_Xray}) is consistent with gas piling up along expanding fronts, which produce lower-luminosity cavities in their wake (Fig.~\ref{fig:x-ray_profiles}).\\

\item Across the TNG50 sample, the bubbles expand with speeds as high as $1000-2000~\KMS$ (about $1-2$ kpc Myr$^{-1}$), but with a great diversity and with larger bubbles typically expanding at slower speeds than smaller ones (Fig.~\ref{fig:velocities}). For bubbles smaller than 20 kpc, the median TNG50 bubble expands radially as fast as $500~\KMS$, but bubbles can expand as slow as 100 $\KMS$. \\

\item Also the inferred bubble ages can be very diverse: bubbles of about 10 kpc in size and that at the time of inspection move at e.g. $1500~\KMS$ ($1000~\KMS$) can be 6 (10) Myr old. However, larger bubbles of $50-60$ kpc height can be as old as $30-100$ Myr or more.\\

\item The high-velocity outflows produce shocks and, in many but not all galaxies, coherent shock fronts at the edges of the dome-like features that are clearly discernible in gas pressure, X-ray or temperature (Fig.~\ref{fig:top30_machnum}). The typical TNG50 bubble develops shocks with $1.8-3.7$ average Mach Numbers (25th-75th percentiles across the bubble sample: Fig.~\ref{fig:velocities}).

\end{itemize}

All in all, the diagnostics uncovered in this paper support a push+shock mechanism for the development of the large-scale bubbles in TNG50 galaxies. In our simulated galaxies, the bubbles are produced by kinetic, wind-like feedback driven by the SMBHs at the galaxy centers (Figs.~\ref{fig:timeevol_1}, \ref{fig:timeevol_2}), as the star formation in the inner regions is instead typically and mostly suppressed (Section~\ref{sec:nosfr}). Episodic and subsequent events of energy injection by the SMBHs inflate single bubbles, in that, typically, TNG50 bubbles appear to be the result of multiple activity bursts that add up and drive them. In our model and in MW/M31-like galaxies, such episodic feedback events may manifest into bubbles every $20-50$ million years, with the SMBHs typically accreting at low Eddington ratios (Fig.~\ref{fig:smbhs}) and with multiple bubble features -- younger than about $150-160$ Myr and of progressively larger size -- often coexisting within the CGM of the same galaxy.

A striking aspect of our findings is their diversity, whereby the same implementation of SMBH feedback -- together with the other ingredients of the underlying galaxy formation model in the full cosmological context -- returns a wide variety of configurations and physical properties of the gas, across different galaxies, within the same galaxy and within the same galaxy across time.

\subsection{Predictions and observational signatures}

We find the following three broader implications from our analysis and numerical model:

\begin{enumerate}

\item Our results from TNG50 suggest that features similar to the eROSITA and Fermi bubbles of our Milky Way could quite plausibly be produced by kinetic winds or small-scale jets related to the activity of the central SMBH, without requiring the latter to be in a quasar, i.e. high-accretion, phase. \\

\item If the mechanism that inflated the bubbles in the Milky Way is episodic as is the case in TNG50, other features of piled up gas in coherent fronts may be present in the CGM of the Galaxy, at larger galactocentric distances.\\

\item According to TNG50, X-ray (and possibly $\gamma$-ray) bubbles similar to or more extended than the ones seen in the Milky Way could be a frequent and rather ubiquitous feature of disk-like galaxies prior to, or on the verge of, being quenched.

\end{enumerate}

Whether SMBH-driven bubbles could be detected in the CGM of Andromeda remains to be determined: our findings motivate further efforts, both theoretical and observational, in the direction of our neighboring M31. In fact, the X-ray luminosities of the bubble galaxies predicted by TNG50 should be observable in the local Universe with the eROSITA and Chandra telescopes with reasonable observing times (Figs.~\ref{fig:x-ray_profiles} and \ref{fig:top30_Xray}). X-ray stacking of many MW/M31-like galaxies may be the avenue to uncover the bubble phenomenology in external galaxies \citep{Truong.2021b}. The angular modulation of the X-ray emission should be detectable in the nearby Universe, despite it being small: in fact, whereas the gas density is lower along the minor axis of galaxies according to our predictions, the gas temperature and metallicity are higher (Fig.~\ref{fig:metals}, \citealt{Peroux.2020}, and \citealt{Truong.2021b}).

As another manifestation of the preferred direction of feedback, we have uncovered in both SDSS data and in the IllustrisTNG simulations that the satellite quenched fractions in stacked groups of galaxies are lower along the minor axis of their central \citep{MartinNavarro.2021}. From our analysis of both observational and simulated data, we conclude that this `quenching-anisotropy signal' is due to the interaction between satellite galaxies and the CGM, the latter modulated by the SMBH activity of the central galaxies. Namely, the `quenching-anisotropy signal' is a population-wide manifestation of SMBH feedback carving lower-density regions in the CGM around the centrals, particularly along the minor axis, similar to the process for individual bubbles in MW/M31-like galaxies, but integrated over cosmic epochs, populations of galaxies, and mass ranges.

\subsection{Future directions}

The development of automated methods for the identification of (simulated) galaxies with bubble-like features would make it possible to expand the sample of analyzed galaxies to higher redshifts, lower and higher masses, and multiple projections and more frequent simulation snapshots. This is in turn will permit us to identify what properties of the gaseous disk and of the gaseous halo are needed for coherent CGM features to develop and in practice to extract theoretically-motivated galaxy -- bubble scaling relations. It will also allow us to determine how bubbles evolve into one another and hence their scaling laws, at least in the models, and if the bubbles achieve, or are causally connected to, star-formation quenching in low-redshift galaxies. 

Whereas we caution against the quantitative over-interpretation of the similarities between the eROSITA/Fermi bubbles and the TNG50 bubbles described in this paper, we notice that more sophisticated forward-modeling analyses of TNG50 galaxies are in principle possible, beginning with more tailored X-ray mocks, predictions for $\gamma$-ray radiation and cosmic ray modeling, as well as e.g. mocks of Faraday rotation measures -- magnetic fields are in fact consistently modeled in the simulation. In particular, models of the $\gamma$-ray emission assuming the two classes of mechanisms advocated for the Fermi bubbles, hadronic and leptonic, are in principle achievable starting from the TNG50 output \citep[e.g. by following the prescriptions of][]{Sarkar.2015}, although assumptions related to the cosmic-ray proton energy density and with models for the energy distribution of relativistic electrons are needed \citep{Marinacci.2018}. On similar veins, extracting the TNG50 predictions for the UV emission and absorption signatures and kinematics in the bubble regions, the OVII to OVIII absorption column density ratios as a probe of the bubble gas temperature, and the transverse vs. vertical velocity of the bubble gas could all provide insightful expectations, also for the more general goal of probing the manifestations of SMBH feedback in external galaxies. 

We anticipate TNG50 to be a particularly useful, and unique, laboratory to build theoretically-motivated expectations for the interplay between SMBH-driven winds and magnetic field properties of the gas in the circumgalactic gas and to understand what role magnetic fields can have in inflating the bubbles.

\section*{Data Availability}
The IllustrisTNG simulations, including the most recent TNG50, are publicly available and accessible at \url{www.tng-project.org/data} \citep{Nelson.2019release}. Data directly related to this publication and its figures are available upon request from the corresponding author or are partially accessible at the IllustrisTNG webpage.

\section*{Acknowledgements}
AP and NT acknowledge support from the Deutsche Forschungsgemeinschaft (DFG, German Research Foundation) -- Project-ID 138713538 -- SFB 881 (``The Milky Way System'', subproject C09). This work is also supported by the DFG under Germany’s Excellence Strategy EXC 2181/1-390900948 (the Heidelberg STRUCTURES Excellence Cluster). DN acknowledges funding from the DFG Deutsche Forschungsgemeinschaft (DFG) through an Emmy Noether Research Group (grant number NE 2441/1-1). The TNG50 simulation was realised with compute time granted by the Gauss Centre for Super-computing (GCS) under the GCS Large-Scale Project GCS-DWAR (2016; PIs Nelson/Pillepich). This research was supported in part by the National Science Foundation under Grant No. NSF PHY-1748958: AP thanks the organizers and participants of the KITP's Halo21 program, particularly Kartick Sarkar and Rongmon Bordoloi, for inspiring discussions and useful inputs.


\bibliographystyle{mnras}
\bibliography{MWBubbles} 



\appendix

\section{More on the physical properties of the gas in the bubbles}
\begin{figure*}
		\includegraphics[trim=0 150 0 0,clip, width=18cm]{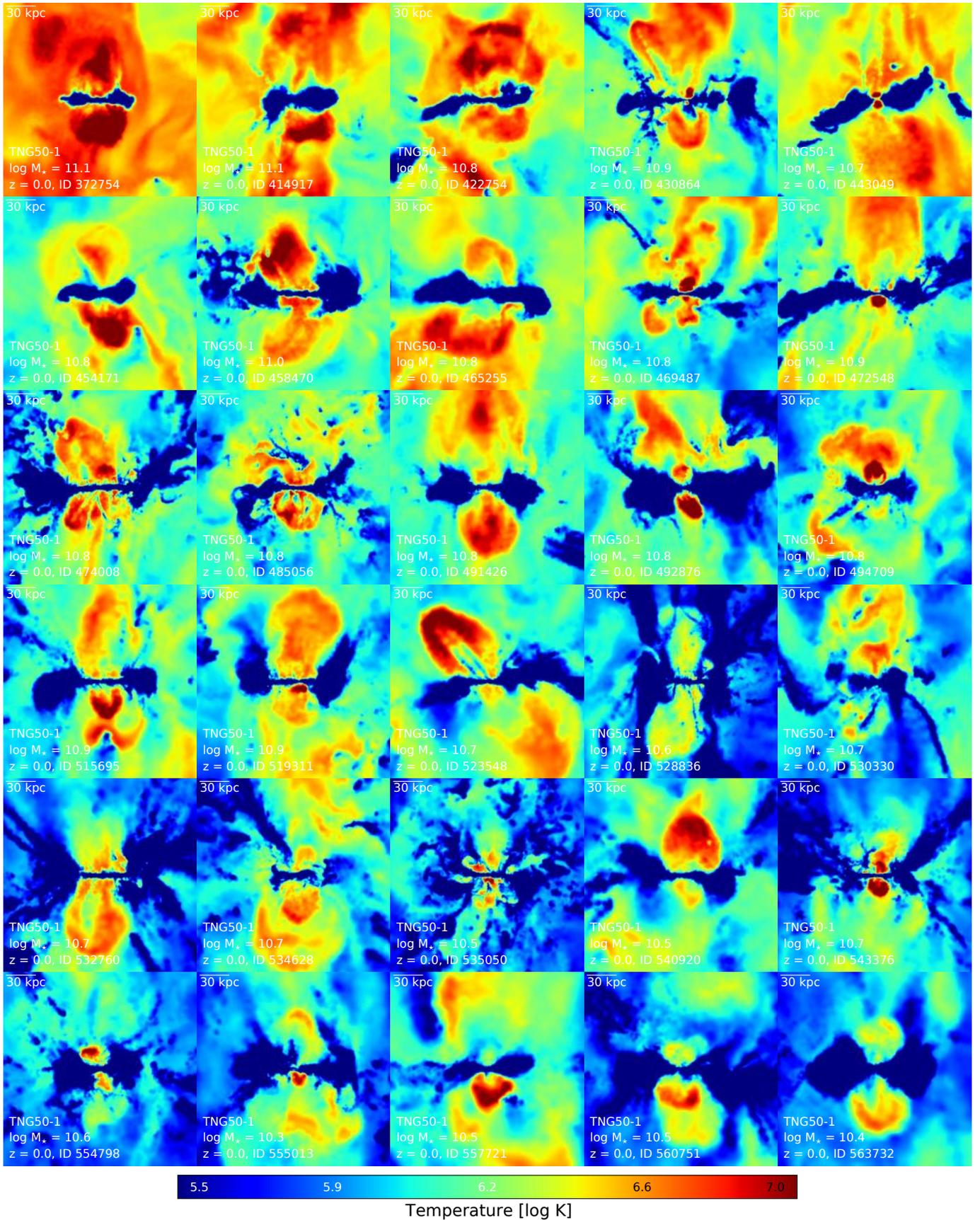}
    \caption{Mass-weighted temperature of the gas of the same galaxies and in the same edge-on projections as in Fig.~\ref{fig:top30_P_gas}.}
    \label{fig:top30_T}
\end{figure*}

\begin{figure*}
		\includegraphics[trim=0 150 0 0,clip, width=18cm]{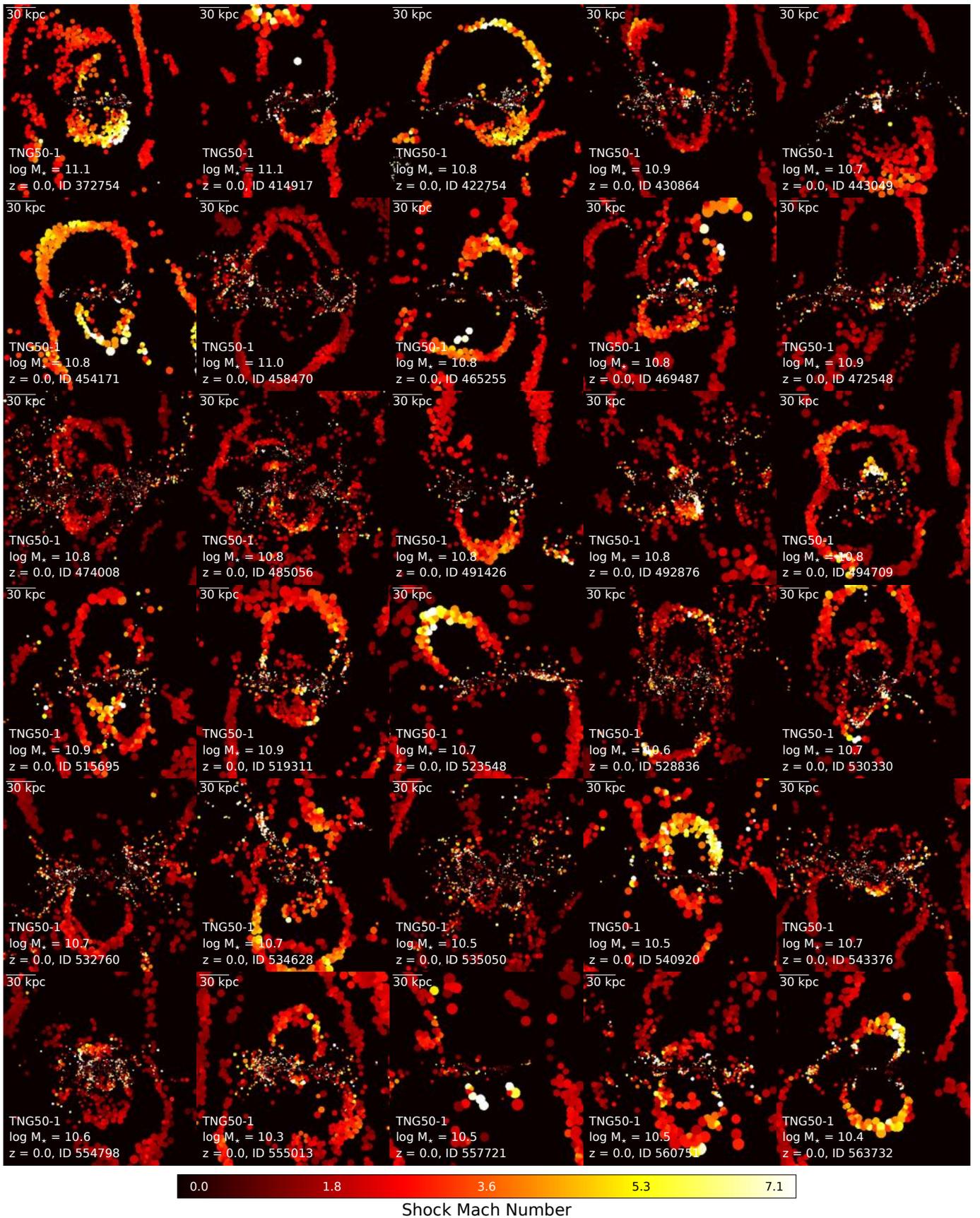}
    \caption{The same galaxies as in Fig.~\ref{fig:top30_P_gas} but now the colors show the mass-weighted averaged Mach Numbers according to our shock finder: here, the averages are obtained by only considering gas cells with shock Mach number larger or equal than 1. Many, but not all, galaxies show clear shock fronts.}
    \label{fig:top30_machnum}
\end{figure*}

In this Appendix, we collect additional maps of the physical properties of the gas above and below the disks of MW/M31-like galaxies from TNG50: in particular, mass-weighted temperature (Fig.~\ref{fig:top30_T}) and  mass-weighted averaged Mach Numbers (Fig.~\ref{fig:top30_machnum}) of the gas in the same TNG50 galaxies and edge-on views of Figs.~\ref{fig:top30_P_gas} and ~\ref{fig:top30_Xray}.

\end{document}